\documentstyle[preprint,aps,prl,psfig]{revtex}
\tightenlines
\begin{document}
\newcommand{\sinf}{\raisebox{-.7ex}{$\stackrel{<}{\sim}$}}
\newcommand{\ssup}{\raisebox{-.7ex}{$\stackrel{>}{\sim}$}}

\ifx\undefined\psfig\def\psfig#1{    }\else\fi
\ifpreprintsty\else
\twocolumn[\hsize\textwidth
\columnwidth\hsize\csname@twocolumnfalse\endcsname       \fi    \draft
\preprint{  }  \title  {Coulomb interaction effects in spin-polarized
transport}
\author  {Irene D'Amico}
\address{Istituto Nazionale per la Fisica della Materia (INFM)}
\address{Institute for Scientific Interchange (ISI), Viale Settimio Severo 65,
I-10133 Torino, Italy}
\author{Giovanni Vignale}
\address{Department  of Physics,   University   of
Missouri, Columbia, Missouri 65211} \date{\today} \maketitle

\begin{abstract} We study the effect of the electron-electron interaction
on the transport of spin polarized currents in metals and doped
semiconductors in the diffusive regime.  In addition to well-known
screening effects, we identify two additional effects, which depend on
many-body correlations and exchange and {\it reduce} the spin diffusion
constant.  The first is the  ``spin Coulomb drag" - an
intrinsic friction mechanism which operates whenever the average velocities of
up-spin and down-spin electrons differ.  The second arises from the decrease in the longitudinal
spin stiffness of an interacting electron gas relative to a noninteracting one.
Both effects are studied in detail for both degenerate and non-degenerate
carriers  in
metals and semiconductors, and various
limiting cases are worked out analytically.  The behavior of the spin diffusion
constant at and below a ferromagnetic transition temperature is also
discussed.
  \end{abstract} \ifpreprintsty\else\vskip1pc]\fi \narrowtext

\section{Introduction}
The theory of spin-polarized transport  is attracting the attention
of the physics community both for its potential applications to
the emerging field of  ``spintronics'' \cite{Prinz} and in relation to the
intriguing field of quantum computation\cite{loss}.  In this context Kikkawa et
al.  (\cite{Awscha1,Awscha5})  generated  much excitement by demonstrating
optical injection and subsequent control of packets of spin polarization in the
conduction band of n-doped GaAs.   Such packets could  in principle be used to
transport information between separate regions of a system.   Unlike  ordinary
electron-hole packets, whose mobility is   limited by strong  scattering in
the
hole component,   these {\it unipolar} electronic spin packets are both
long-lived  (with lifetime $\tau_s \stackrel{>}{\sim} 10 ~ns$) and highly
mobile  \cite{Awscha5}.

 From the theoretical point of view,  most recent work on spin transport
phenomena is based  on an independent electron model, where, in general, the
coupling between different spin channels is  completely neglected.
 Flatt\`e and Byers \cite{Flatte} have recently discussed the behavior of
spin packets in semiconductors in the framework of the Hartree
approximation,   where  the Coulomb interaction is taken into account only
through the imposition of a charge neutrality constraint.  This
constraint establishes an electrostatic coupling between different spin
channels.    They neglect, however,  all further many-body effects.

 In this paper we go beyond the treatment by Flatt\'e and Byers by considering
more subtle correlation (spin-drag) and exchange (spin-softening) effects.
Our objective is the  derivation of  drift-diffusion equations for
spin packets in a full many-body context, i.e., allowing for
correlation between different spin channels.   These equations contain the
spin-packet mobility and the diffusion constant as key parameters.  We show
that both exchange interactions and the drag effect between different spin
populations \cite{spindrag}, concur in reducing the value of the diffusion
constant by a sizable amount, leaving at the same time the mobility basically
unaffected. The microscopic quantities responsible for this effect are the {\it
longitudinal} spin stiffness (the second derivative of the free energy with
respect to magnetization,  not to be confused with the transverse spin
stiffness
of the ferromagnetic state) and the spin-drag transresistivity.   We discuss in
detail the behavior of these quantities in   various regimes and  show how
important the Coulomb effects are when dealing with doped semiconductors
(see Sec. \ref{secsd}).

This paper is organized as follows:

In Section II we review the basic ideas underlying the spin-resolved drift
diffusion equations and make use of the Landau transport
equation to elucidate the structure  of the homogeneous spin resistivity
matrix.   In particular, we show that the off-diagonal element of
the spin-resistivity matrix  (the spin trans-resistivity) is almost exclusively
controlled by the Coulomb interaction:  the contribution from spin-flip
scattering, while finite,  is utterly negligible  for short-range scatterers.

In Section  III we present the calculation of the Coulomb contribution to the
spin trans-resistivity.  The theory of Ref.~\cite{spindrag} is extended in various
directions.  First we study the spin drag effect as a function of temperature
going from the degenerate regime (which is appropriate for  ordinary metals) to
the nondegenerate regime, which is appropriate for low-density/high
-temperature doped semiconductors.  Then, we calculate the spin drag in the
``mixed" case in which one spin component is degenerate while the other is
nondegenerate:  this is relevant to situations in which a strong spin
polarization exists.

Section IV is devoted to a description of the behavior of the longitudinal spin
stiffness of the homogeneous electron gas as a function of density and
temperature.

In Section V we present a detailed derivation of the drift-diffusion equation
for a macroscopic spin packet, and give explicit expressions for the mobility
and diffusion constant in terms of microscopic quantities such as the spin
stiffness and the spin transresistivity.

Section VI  suggests experiments aimed at directly measuring  the spin
transresistivity  and the spin diffusion constant in metals or semiconductors.

Electron gas theory predicts that, at sufficiently low temperatures and
densities,
the homogeneous electron gas undergoes a ferromagnetic transition
\cite{Ceperley,Ortiz}.  Such a transition could be in principle 
 observed in a doped
semiconductor .  Section VII  examines the behavior of spin diffusion constant
in the paramagnetic phase and
how  the occurrence of ferromagnetism as a second order phase
transition would affect it.  Our calculations show
that
the diffusion constant vanishes at the transition temperature and increases as
the system  becomes fully spin polarized with decreasing temperature.

\section{General Theory I} \subsection {The drift-diffusion equation} The
theory of diffusive transport in metals and semiconductors is based on the
assumption of a local linear relationship between the current densities of
up and down spin electrons and the gradient of the local electro-chemical
potentials:  \begin{equation} \label{eq1} -e\vec J_\alpha (\vec r)=
-\sum_\beta \sigma_{\alpha \beta} (\vec r) \vec \nabla\psi_\beta (\vec r).
\end{equation}
 Here $e$ is the absolute value of the electron charge, $\vec J_\alpha
(\vec r)$ ($\alpha = \uparrow$ or $\downarrow$) are {\it number currents}
(\cite{footnote0}) and the
{\it electro-chemical} potentials $\psi_\alpha(r)$ are given by the
electrostatic potential $\phi(\vec r)$ plus the local chemical potential,
which can be spin-dependent,
 \begin{equation} \label{electrochemicalpotential} \psi_\alpha (\vec r) =
\phi(\vec r) - {1 \over e} {\partial f (n_{\uparrow}, n_{\downarrow},T)
\over \partial n_\alpha}. \end{equation}
Here $f (n_{\uparrow}, n_{\downarrow},T)$ is the free energy per unit volume
of a homogeneous interacting electron gas evaluated at the local spin densities
$n_{\alpha} (\vec r)$ and uniform temperature $T$ \cite{footnote1};
 $\sigma_{\alpha \beta} \equiv  \lim_{\omega \to 0} \sigma_{\alpha
\beta}(q=0, \omega)$ is the homogeneous conductivity matrix of the electron
gas, whose structure, in the presence of interactions, will be elucidated
below.

Substituting (\ref{electrochemicalpotential}) in (\ref{eq1}) and writing
\begin{equation}
\vec \nabla{\partial f (n_{\uparrow}, n_{\downarrow},T)
\over \partial n_\alpha} = \sum_\beta {\partial^2 f (n_{\uparrow},
n_{\downarrow},T) \over \partial n_\alpha \partial n_\beta}  \vec \nabla n
_\beta,
\end{equation}
we obtain
 \begin{equation} \label{driftdiffusion} e \vec J_\alpha (\vec r)  =
\sum_{\beta} \left ( \sigma_{\alpha \beta} (\vec r) \vec \nabla \phi (\vec
r) - e D_{\alpha \beta} (\vec r) \vec \nabla n_{\beta} (\vec r) \right )
\end {equation} where the diffusion matrix $D_{\alpha \beta}$ is given by
\begin{equation} \label{diffusionmatrix} e^2 D_{\alpha \beta} =
\sum_\gamma \sigma_{\alpha \gamma} S_{\gamma \beta} \end{equation} and
\begin{equation} \label{spinstiffness} S_{\alpha \beta} = {\partial^2 f
(n_{\uparrow}, n_{\downarrow},T) \over \partial n_{\alpha} \partial
n_{\beta}} \end{equation} is the static longitudinal spin-stiffness matrix
- the
inverse of the spin susceptibility matrix $\chi_{\alpha \beta}$.
Eq.~(\ref{diffusionmatrix})  is the well known Einstein relation between
the mobility and the diffusion constant, generalized here to the case of
spin polarized transport.  The first term of Eq.~(\ref{driftdiffusion})
is the drift current associate with the electrostatic field, the second is
the diffusion current associated with the gradient of the electronic
densities.  These two terms cancel out exactly in a situation of local
equilibrium, due to the Einstein relation  and the fact that
$[\chi^{-1}]_{\alpha \beta}=S_{\alpha \beta}$.

On a formal level the main effect of the Coulomb interaction is the
appearance of non-vanishing {\it off-diagonal elements} of the
conductivity and spin-stiffness matrices.  $\sigma_{\uparrow
\downarrow} \neq 0$ implies that an electric field acting only on the
up-spin electrons must necessarily drag along a current of down-spin
electrons.  Conversely, a current of up-spin electrons $J_\uparrow \neq 0$
flowing against a background of stationary down-spin electrons
($J_\downarrow =0$)  will necessarily induce a gradient of spin-down
electro-chemical potential $\vec{E}_\downarrow = - \nabla \psi_\downarrow
= e\rho_{\downarrow \uparrow} \vec{J}_\uparrow$, where $\rho_{\alpha
\beta}$ is the resistivity matrix, inverse to $\sigma_{\alpha \beta}$.  We
shall see later how one can make use of these effects for a direct
experimental measure of the spin Coulomb drag.

The other noteworthy feature is  $S_{\uparrow \downarrow} \neq 0$:
this means that the chemical potential of up-spins $\partial f
(n_{\uparrow}, n_{\downarrow},T)  /\partial n_{\uparrow}$ is a function of
{\it both} up and down spin densities. Thus,  a disturbance acting on one of
the two spin populations will affect   the other through Coulomb
correlation.

\subsection {Structure of the resistivity matrix} \label{secstrusd}
Although the homogeneous resistivity matrix can be calculated from
first-principle Kubo formulas and/or transport equations its general structure
(including the off-diagonal terms, due mostly to the Coulomb interaction)  is
best understood at the phenomenological level.  We first  present the
phenomenological picture and then justify it from a more formal
consideration of the Landau transport equation.

 Let ${\vec E}_\uparrow(t)$ and ${\vec E}_\downarrow(t)$ be uniform
effective electric fields, ${\vec E}_\alpha=-\nabla \psi_\alpha$, that couple 
to up- and down-spins respectively. 
We restrict ourselves to the linear response regime.  If ${\vec v}_\alpha$ is
the velocity of the center of mass of electrons of spin $\alpha$, and
$N_\alpha$ the number of such electrons, then the equation of motion for
${\vec v}_\alpha$ has the form
\begin{equation} m^*N_{\alpha} \dot{{\vec
v}}_\alpha = -eN_{\alpha} {\vec E}_\alpha + {\vec F}^C_{\alpha
\bar{\alpha}} -\frac{m^*}{\tau_\alpha} N_{\alpha} {\vec v}_\alpha
+\frac{m^*}{\tau^\prime_{\alpha}} N_{\bar \alpha} {\vec
v}_{\bar \alpha},  \label{eqmot}
\end{equation}
where $m^*$ is the effective mass of the carriers  and $\bar{\alpha}\equiv
-\alpha$.

Let us examine the meaning of the various terms in (\ref{eqmot}).

The first term on the right hand side is the net force exerted by the
electric field
on spin-$\alpha$ electrons.

The second term,  ${\vec F}^C_{\alpha\bar{\alpha}}$, is the Coulomb force
exerted by spins of the {\it opposite} orientation $\bar \alpha$ on
spin-$\alpha$
electrons.   Notice that the  net force exerted by spins of the same
orientation
vanishes by virtue of Newton's third law.  For exactly the same reason we must
have  ${\vec F}^C_{\alpha \bar{\alpha}}=-
{\vec F}^C_{{\bar \alpha}\alpha}$, and by Galilean invariance this force can
only depend on the relative velocity of the two components. Hence, in the
linear approximation, we can write
\begin{equation} {\vec F}^C_{\alpha\bar{\alpha}}= - \gamma m^* N_\alpha
\frac{n_{\bar\alpha}}{n}({\vec v}_\alpha-{\vec v}_{\bar{\alpha}}),
\label{gamma1} \end{equation} where $n=n_\uparrow + n_\downarrow$ is
the total density. Eq.~(\ref{gamma1}) defines the {\it spin drag coefficient}
$\gamma$.

The third term on the right hand side  of (\ref{eqmot}) combines two distinct
physical effects.  One is the net force exerted on  spin-$\alpha$
electrons by electron-impurity collisions  that {\it do not} flip
the spin of the incoming electrons.   The other is the  rate at
which momentum is lost to  the $\alpha$ component as a result of
electron-impurity collisions that  flip the spin from $\alpha$ to $\bar
\alpha$.
Accordingly, the momentum relaxation rate $\tau^{-1}_\alpha$ is written as
the sum of non-spin-flip and spin-flip contributions:  $\tau_\alpha^{-1} =
\tau_{nf,\alpha}^{-1}+\tau_{sf,\alpha}^{-1}$.    Spin-flip times
have been found to be very long both in metals ($\tau_{sf, \alpha} \sim
10^{-1} ns$\cite{Silsbee}) and in semiconductors ($\tau_{sf, \alpha}\sim 10
ns$\cite{Awscha1,Awscha5}),
  whereas  non-spin-flip times are usually  much shorter  (of
the order $10^{-3} - 10^{-4}~ns$).    Thus, for most practical purposes,
$\tau_\alpha^{-1}\approx \tau_{nf,\alpha}^{-1}$.

Finally, the last term on the right hand side  of (\ref{eqmot}) represents the
rate at which momentum is fed into the $\alpha$ component
by electron-impurity collisions that  flip the spin from $\bar \alpha$ to
$\alpha$.   Because electrons emerging from collisions have a randomized
momentum distribution   with nearly zero average, we expect  
$(\tau^\prime_{\alpha})^{-1}$ to be  smaller than even the already small spin-flip
rate $\tau_{sf, \alpha}^{-1}$.  Indeed, we shall see in the next
section that  $(\tau^\prime_\alpha)^{-1}$ vanishes for
short-range scatterers in the Born approximation.

Fourier-transforming the equation of motion ~(\ref{eqmot})  for the
current density ${\vec j}_\alpha(\omega)= -e n_\alpha{\vec
v}_\alpha(\omega)$  we find
\begin{eqnarray} \label{eqmotf2}
i\omega{\vec j}_\alpha(\omega) = &-& \frac{n_\alpha e^2}{m^*}{\vec E}_\alpha
(\omega)  +\left ( \frac{n_{\bar\alpha}}{n} \gamma + \frac {1}
{\tau_\alpha} \right ){\vec j}_\alpha (\omega) \nonumber
\\ &-&\left  ( \frac{n_\alpha}{n} \gamma   +   \frac {1}
{\tau^\prime_{\alpha}}  \right ) {\vec j}_{\bar{\alpha}}
(\omega).
\end{eqnarray}
The resistivity matrix $\rho_{\alpha\alpha'}$ is
defined as the coefficient of proportionality between the electric field
and the
current, i.e. 
 ${\vec E}_\alpha = \sum_{\alpha'}\rho_{\alpha\alpha'} {\vec
j}_{\alpha'}$.  A quick comparison between this definition and
Eq.~(\ref{eqmotf2}) shows that
 the
complete form of the resistivity matrix $\rho_{\alpha \beta}$ is
\begin{equation} \label{resistivity} \rho = \left(
\begin{array}{cc}-i\omega{ m^*\over e^2n_\uparrow}+{m^* \over n_\uparrow
e^2 \tau_\uparrow}+{n_\downarrow \over n_\uparrow} {m^* \over ne^2}\gamma
 &  -  {m^* \over n_\uparrow
e^2 \tau^\prime_\uparrow}-{m^* \over ne^2}\gamma \\
 -  {m^* \over n_\downarrow 
e^2 \tau^\prime_\downarrow}-{m^* \over ne^2}\gamma &
-i\omega{ m^*\over e^2n_\downarrow}+{m^* \over n_\downarrow e^2
\tau_\downarrow}+{n_\uparrow \over n_\downarrow} {m^* \over ne^2}\gamma
 \end{array} \right ).  \end{equation}
Notice that this matrix is symmetric, due to the relation $1/n_\downarrow
\tau^\prime_\downarrow =  1/n_\uparrow \tau^\prime_\uparrow$, which will
be proved in the next section.

Due to the extreme smallness of  the spin-flip   rates
$1 /\tau^\prime_\alpha$, the off-diagonal resistivity (also referred to  as
``spin-transresistivity") is controlled almost entirely by the Coulomb
interaction term, i.e. we can safely assume 
\begin{equation}  \label{gamma}
\rho_{\uparrow \downarrow}= -{m^* \over ne^2}\gamma
 \end{equation} and $\gamma$ is directly proportional to the
spin trans-resistivity.
  But then, Galilean
invariance and Newton's third law demand that the same $\rho_{\uparrow
\downarrow}$ appear also as a correction to the ordinary diagonal resistivity.
This effect is quite distinct from the ``trivial" renormalizations due to the
electronic screening of these interactions.

\subsection {Derivation from Landau transport equation}

In this section we provide a microscopic justification of the
phenomenological Eq.~(\ref{eqmot}) and give explicit expressions for the
electron-impurity relaxation rates.

We start from the linearized transport equation for the quasiparticle
distribution function  in Landau theory of Fermi liquid \cite{Nozieres}.
This is simply the classical Boltzmann  equation for quasiparticles in a
self-consistent field described  by Landau parameters.   In the homogeneous
case  the self-consistent field coincides with the classical electric field
and the
distribution function $f_\alpha (\vec k ,t)$ obeys the well-known kinetic
equation
\begin{equation} \label{boltzmann}
{\partial  f_\alpha (\vec k ,t) \over \partial t} - e \vec E_\alpha \cdot
{\partial  f^{(0)}_\alpha (\vec k  ) \over \partial \vec k}  = \left (
{\partial  f_\alpha (\vec k ,t) \over \partial t} \right )_{coll},
\end{equation}
where $f^{(0)}_\alpha (\vec k  )$ is the equilibrium distribution function.

The collision term  $\left ({\partial  f_\alpha (\vec k ,t) / \partial
t} \right
)_{coll}$  has contributions from the Coulomb interaction as well as
spin-flip and non spin-flip electron-impurity interactions.  The various
contributions are listed below

\begin{itemize}
\item {Coulomb collisions}
\begin{eqnarray} \label{Coulombcoll}
\left ({\partial  f_\alpha (\vec k ,t) \over \partial t} \right )^C_{coll}
&=& -
\sum_{\vec p \vec k' \vec p',  \beta \alpha' \beta'}
W^C(\vec k \alpha, \vec p \beta;\vec k' \alpha', \vec p' \beta')
\delta_{\vec k +
\vec p, \vec k' + \vec p'} \delta_{\alpha + \beta, \alpha' + \beta'}
\nonumber \\
&& \left \{ f_\alpha(\vec k)f_\beta(\vec p) [1 - f_{\alpha'} (\vec k') ] [1
-f_{\beta'} (\vec p')] - [1-f_\alpha(\vec k)][1-f_\beta(\vec p)]  f_{\alpha'}
 (\vec
k') f_{\beta'} (\vec p') \right \} \nonumber \\
&& \delta (\epsilon_{k \alpha} + \epsilon_{p \beta} -\epsilon_{k',\beta'} -
\epsilon_{p' \beta'}),
\end{eqnarray}
where $ W^C(\vec k \alpha, \vec p \beta;\vec k' \alpha', \vec p' \beta')$
is the
probability of the  Coulomb scattering process $\vec k \alpha, \vec p \beta \to
\vec k' \alpha', \vec p' \beta'$ and $\epsilon_{k
\alpha}$ is the energy of a particle of momentum $\vec k$ and  spin
$\alpha$ relative to the chemical potential.  The conservations of  momentum,
energy , and spin are explicitly displayed.

\item{Non-spin-flip electron-impurity collisions}
\begin{equation}
\label{nonflipcoll}
\left ( {\partial  f_\alpha (\vec k ,t) \over \partial t} \right
)^{nf}_{coll} =
- \sum_{\vec k'} W^{nf}(\vec k \alpha, \vec k' \alpha)[ f_\alpha(\vec k) -
f_\alpha(\vec k')]  \delta(\epsilon_{k \alpha} - \epsilon_{k' \alpha}),
\end{equation}
where $W^{nf}(\vec k \alpha, \vec k' \alpha)$ is the probability of the
non-spin flip (nf) scattering process $ \vec k \alpha \to \vec k' \alpha$.

\item{ Spin-flip electron-impurity collisions}
\begin{equation} \label{flipcoll}
\left ({\partial  f_\alpha (\vec k ,t) \over \partial t} \right )^{sf}_{coll} =
- \sum_{\vec k'} W^{sf}(\vec k \alpha, \vec k'  \bar \alpha)[ f_\alpha(\vec
k) -
f_{\bar \alpha}(\vec k')]  \delta(\epsilon_{k \alpha} - \epsilon_{k' \bar
\alpha}), \end{equation}
where $W^{sf}(\vec k \alpha, \vec k'  \bar \alpha)$ is the probability of the
spin flip (sf) scattering process $ \vec k \alpha \to \vec k' \bar
\alpha$.
\end{itemize}

In order to obtain a closed equation of motion for the currents, such
as Eq.~(\ref{eqmotf2}),  we must multiply both sides of
Eq.~(\ref{boltzmann}) by $ -e \vec k /m^*$ , sum over $\vec k$, and then
 express the integrated collision term
\begin{equation} \label{totalforce}
\vec F_\alpha \equiv \sum_{\vec k} \vec k
\left ({\partial  f_\alpha (\vec k ,t) \over \partial t} \right )_{coll}
\end{equation}
back in terms of the currents.  Of course, this cannot be done rigorously,  but
for an isotropic system slightly perturbed from equilibrium one can assume
\cite{Mermin}  that the distribution function of the state with currents $\vec
j_\alpha = - e n_\alpha \vec v_\alpha$ is given by
\begin{equation} \label{ansatz}
f_\alpha (\vec k ,t)  =   f^{(0)}_\alpha (\epsilon_{k \alpha} ) - {\partial
f^{(0)}_\alpha (\epsilon_{k \alpha} )  \over \partial   \epsilon_{k \alpha} }
\vec v_\alpha (t) \cdot \vec k.
\end{equation}
Substituting this into Eqs.~(\ref{Coulombcoll}), (\ref{nonflipcoll}), and
(\ref{flipcoll}), and linearizing with respect to the currents wherever
needed,
we arrive, after tedious but straightforward manipulations, at the
desired equation of motion (\ref{eqmotf2}), with the following expressions
for the various relaxation times:
\begin{eqnarray} \label{taucoulomb}
\gamma  &=& {n \over m^* N_\alpha n_{\bar \alpha}} \sum_{\vec k \vec p
\vec k' \vec p'} {(\vec k - \vec k')^2 \over 2 d k_B T}
W^C(\vec k \alpha, \vec p \bar \alpha;\vec k' \alpha, \vec p' \bar \alpha)
\delta_{\vec k + \vec p, \vec k' + \vec p'} \delta (\epsilon_{k \alpha} +
\epsilon_{p \bar \alpha} -\epsilon_{k',\alpha} - \epsilon_{p' \bar \alpha})
\nonumber \\ &&
 f^{(0)}_\alpha (\epsilon_{k \alpha} ) f^{(0)}_{\bar \alpha} (\epsilon_{p
\bar \alpha} )  f^{(0)}_\alpha (-\epsilon_{k' \alpha} )
f^{(0)}_{\bar \alpha} (-\epsilon_{p'  \bar \alpha} ),
\end{eqnarray}
where $d$ is the number of spatial dimensions;
\begin{equation}
\label{taunf} {1 \over \tau_{nf, \alpha}} = - \sum_{\vec k} {\partial
f^{(0)}_\alpha (\epsilon_{k \alpha} )  \over \partial   \epsilon_{k \alpha} }
{k^2 \over n_\alpha d} \sum_{\vec k'}
W^{nf}(\vec k \alpha, \vec k' \alpha) (1 - \hat k \cdot \hat k')
\delta(\epsilon_{k \alpha} - \epsilon_{k' \alpha}),
\end{equation}
where $\hat k$ and $\hat k'$ are unit vectors in the directions of $\vec k$ and
$\vec k'$; 
\begin{equation} \label{tausf}
{1 \over \tau_{sf, \alpha}} = - \sum_{\vec k} {\partial
f^{(0)}_\alpha (\epsilon_{k \alpha} )  \over \partial   \epsilon_{k \alpha} }
{k^2 \over n_\alpha d} \sum_{\vec k'}
W^{sf}(\vec k \alpha, \vec k'  \bar \alpha)
\delta(\epsilon_{k \alpha} - \epsilon_{k'  \bar \alpha}),
\end{equation}

\begin{equation} \label{tauprime}
{1 \over \tau^\prime_\alpha} = - {1 \over n_{\bar \alpha} d}  \sum_{\vec k}
{\partial f^{(0)}_\alpha (\epsilon_{k \alpha} )  \over \partial   \epsilon_{k
\alpha} } \sum_{\vec k'}
W^{sf}(\vec k \alpha, \vec k'  \bar \alpha) \vec  k \cdot \vec k'
\delta(\epsilon_{k \alpha} - \epsilon_{k'  \bar \alpha}).
\end{equation}

Notice that $1/\tau_{sf,\alpha}$ and $1 /\tau^\prime_\alpha$ arise, respectively, from the
first and
the second term on the right side of Eq.~(\ref{flipcoll}).  The key difference
between these two relaxation rates is that the expression for the latter
involves
an angular average of the scattering probability with weight factor $\hat k
\cdot \hat k' = \cos (\theta)$.  This average vanishes in the Born
approximation for short-range scatterers, since the scattering probability
becomes isotropic (independent of $\vec k$ and $\vec k'$) in this special case.
Quite generally, one can expect $1/\tau^\prime_\alpha$ to be much smaller than
$1/\tau_{sf,\alpha}$ in agreement with the qualitative arguments given in the
previous section.  This means that the spin transresistivity is almost entirely
a  Coulomb interaction effect, and therefore its measurement can shed light on
the nature of the Coulomb correlation between up- and down-spin electrons.
This is one of the main points we wanted to make in this section.

Finally, notice that Eq.~(\ref{tauprime}) implies the identity
\begin{equation} \label{tauprimesymmetry}
{1   \over  n_\downarrow
\tau^\prime_\downarrow } =  {1 \over n_\uparrow \tau^\prime_\uparrow}
\end{equation}
which guarantees the symmetry of the resistivity matrix 
Eq.~(\ref{resistivity}).

\section{Calculation of the spin transresistivity} \label{secsd} The
 theory of the spin transresistivity has been worked out in
\cite{spindrag}.  This theory closely parallels the theory of the ordinary
Coulomb drag between parallel two-dimensional electron or hole-gas layers
\cite{Rojo} but  differs in some important details, as the fact that 
 electrons of opposite spin interact with the {\it same} set of
impurities, so that certain electron-impurity terms which appear in the
Kubo formulation of the transresistivity do not vanish upon disorder
averaging. Fortunately, it  turned out that these terms cancel out exactly
at low
frequency ($\omega <<E_F$) and to leading order in the electron-electron and
electron-impurity interactions \cite{spindrag}.

In this section we first review for completeness the derivation of Ref.
\cite{spindrag}, and then present an RPA calculation of $\rho_{\uparrow
\downarrow}$ {\it at finite temperature} \cite{footnote2}.  At variance
with the
calculation of Ref. \cite{spindrag} we present our results not only in the
low temperature limit (which is relevant to metals and where $\gamma \sim
T^2$), but also in the non-degenerate $k_BT>>E_F$ and quasi-degenerate
$k_BT \sim E_F$ regimes (which are relevant to doped semiconductors),
where $k_B$ is the Boltzmann constant and
$E_F=\hbar^2(3\pi^2n)^{2/3}/2m^*$ is the Fermi energy\cite{drag_he}.

We start from the Kubo formula \cite{Kubo} for the uniform conductivity
matrix 
\begin{equation} \label{Kuboconductivity} \sigma_{\alpha, \alpha'}
(\omega)= -{1 \over i \omega} {e^2 \over m^*} \left( n_\alpha \delta_{\alpha,
\alpha'} + {\langle \langle {\vec P}_\alpha ; {\vec P}_{\alpha'} \rangle
\rangle_\omega \over m^*} \right), 
\end {equation} where $\langle \langle A;B
\rangle \rangle_\omega$ represents, as usual \cite{Kubo}, the retarded
response function for the expectation value of $A$ under the action of a
field that couples linearly to $B$.
 We assume to be in the diffusive, weak scattering regime characterized by
$\hbar/\tau_D<<k_BT$,
where $\tau_D=(n_\uparrow/n)\tau_{\uparrow}
+ (n_\downarrow/n)\tau_{\downarrow}$.  Because of the ``high"
temperature, weak localization effects   are negligible.  In this regime the
resistivity is essentially independent of frequency for $\omega <<
k_BT/\hbar$.  It is therefore legitimate to take the limit of weak
electron-impurity and electron-electron scattering {\it before} taking the
limit
of $\omega \to 0$\cite{Goetze}.  When the limits are carried out in this order
the $P_\alpha$'s are almost constants of motion and therefore the second term
in the square bracket of Eq.~(\ref{Kuboconductivity}) is a small correction to
the first.   Inverting Eq.~(\ref{Kuboconductivity}) to first order in
$\langle \langle {\vec P}_\alpha; {\vec P}_{\alpha'} \rangle \rangle_\omega$
and selecting the $\uparrow \downarrow$ matrix element we obtain
\begin{equation} \label{transresistance1} \rho_{\uparrow \downarrow}
(\omega) = {i \omega \over e^2}
 {\langle \langle {\vec P}_\uparrow ; {\vec P}_\downarrow \rangle
\rangle_\omega \over n_\uparrow n_\downarrow}.  \end {equation} It is
convenient to recast this equation in a form that emphasizes the
importance of the non conservation of $P_\uparrow$ and $P_\downarrow$.  To
this end we make use twice of the general equation of motion
\begin{equation} \langle \langle A;B \rangle \rangle_\omega=\frac{1}{
\omega}(\langle[A,B]\rangle +i\langle \langle \dot{A};B \rangle \rangle
_\omega), \end{equation} where $\dot{A} \equiv -i [A,H]$ is the time
derivative of the operator $A$, and $\langle .. \rangle$ denotes the
thermal average.  Thus, Eq.~(\ref{transresistance1}) can be rewritten as
\begin {equation} \label{transresistance2} \rho_{\uparrow \downarrow}
(\omega) = {i \over e^2 n_\uparrow n_\downarrow} {\langle \langle
\dot{{\vec P}}_\uparrow ; \dot{{\vec P}}_\downarrow \rangle \rangle_\omega
+ i\langle[\dot{{\vec P}}_\uparrow,{\vec P}_\downarrow] \rangle \over
\omega}.  \end {equation} The commutator term controls the high frequency
behavior of $\rho_{\uparrow \downarrow} (\omega)$ and can be expressed
 in terms of ground-state properties \cite{Goodman}.
  This term however gives a purely imaginary contribution to the
trans-resistivity.  Our present interest is in the real part of the
trans-resistivity, which is controlled by the imaginary part of the
force-force response function.

 The force operator is given by \begin{equation} \label{force} \dot{{\vec
P}_\alpha}= -\frac{i}{V}\sum_{\vec q}{\vec q}v_q \rho_{{\vec
q}\bar{\alpha}} \rho_{-{\vec q}\alpha} -\frac{i}{V}\sum_{\vec q}{\vec
q}v^{e-i}_q\rho^i_{{\vec q}}\rho_{-{\vec q} {\alpha}}, \end{equation}
where $v_{ q} = 4 \pi e^2/q^2\epsilon$ is the Fourier transform of the
Coulomb interaction, $\epsilon$ the dielectric constant of the material,
$v^{e-i}_q$ is the Fourier transform of the electron-impurity interaction,
$\rho_{{\vec q}\alpha}$ is the electronic spin density fluctuation
operator, $\rho^i_{{\vec q}}$ is the Fourier transform of the impurity
density (a number), and $V$ is the volume of the system.

 As explicitly shown in \cite{spindrag}, the contribution of correlated
impurity scattering to the trans-resistivity,
 in the low frequency $\hbar\omega <<E_F$ limit and to leading order in
the electron-electron and electron-impurity interactions,
vanishes since
 the Coulomb-impurity term {\it exactly} cancels the impurity-impurity
contribution. Thus,in this limit, the real part of the spin
trans-resistivity takes the form 
\begin{eqnarray}
\label{realtransresistance} Re \rho_{\uparrow \downarrow}
(\omega)&=&\frac{1}{n_\uparrow n_{\downarrow}e^2\omega V^2}\sum_{\vec q
\vec q'}\frac{\vec q \cdot \vec q'}{3}
 v_{q}v_{ q'}\cdot \nonumber \\ &~&Im \langle
\langle \rho_{-{\vec q}\uparrow}\rho_{{\vec q}\downarrow};\rho_{{\vec
q'}\uparrow}\rho_{-{\vec q'}\downarrow} \rangle \rangle_\omega.
\label{gamma2} \end{eqnarray}

We proved the cancellation of correlated impurity scattering effects
within the frame of the Drude-Boltzmann theory defined by
Eq.~(\ref{transresistance1}) which is the result of interchanging the
natural order of the $\omega \to 0$ limit and the weak scattering limit.
This approach is only adequate in the ``classical" regime $k_B T >>
\hbar/\tau_D$.  A more sophisticated treatment of quantum effects in
correlated impurity scattering \cite{dis} suggests that the spin drag
 would be {\it even larger} than predicted by the present theory at
temperatures $k_BT << \hbar /\tau_D$.  The temperature range in which
these quantum corrections are important shrinks to zero in the limit of
weak impurity scattering.

 We have calculated the four point response function \\
$\chi_{4\rho}({\vec q},{\vec q'},\omega) \equiv \langle \langle
\rho_{-{\vec q}\uparrow}\rho_{{\vec q}\downarrow};\rho_{{\vec
q'}\uparrow}\rho_{-{\vec q'}\downarrow} \rangle \rangle_\omega$ at finite
temperature in a generalized Random Phase Approximation (RPA).
  Because of its infinite
range, the Coulomb interaction must be treated to infinite order, even
when weak.  The sum of the RPA diagrams\cite{spindrag}
 has been evaluated by standard
methods \cite{Mahan} with the following result:  \begin{eqnarray} &~& Re
\rho_{\uparrow\downarrow}(\omega,T) = \frac{1}{n_\uparrow n_\downarrow
e^2}{1\over V}\sum_{\vec q }\frac{q^2}{3}
v_{q}^2\cdot\frac{(e^{-\beta\omega}-1)}{\omega}\nonumber \\ &~&
\int_{-\infty}^\infty \frac{d\omega '}{\pi}
\frac{[\chi''_{\uparrow\uparrow}(q,\omega ')
\chi''_{\downarrow\downarrow}(q,\omega-\omega ')  -
\chi''_{\uparrow\downarrow}(q,\omega ')
\chi''_{\downarrow\uparrow}(q,\omega-\omega ')]}{(e^{-\beta\omega
'}-1)(e^{-\beta(\omega-\omega ')}-1)}.  \label{rho} \end{eqnarray}
Here $\beta=1/k_B T$, $\chi''_{\alpha \alpha'}(q,\omega)$ is the imaginary
part of the RPA spin-resolved density-density response function, which is
related to the noninteracting response function $\chi_{0\alpha}
(q,\omega)$ as follows \begin{equation} \label {chirpa} [\chi^{-1}(q,
\omega)]_{\alpha \alpha'} = [\chi_{0 \alpha}]^{-1} (q,\omega)
\delta_{\alpha \alpha'} - v_{q}.  \end{equation}

It is possible to show by simple but tedious algebraic calculations that
Eq.~(\ref{rho}) for the spin trans-resistivity
$\rho_{\uparrow\downarrow}(\omega,T)$ reduces, in the case of finite
temperature and $\omega=0$, to the well known result of memory function
and diagrammatic theories for the conventional Coulomb drag
\cite{recenttheory}, \cite{Forster} \begin{equation}\label{rhoom=0} Re
\rho_{\uparrow\downarrow}(0,T) = \frac{\beta}{n_\uparrow
n_\downarrow e^2}{1\over V}\sum_{\vec q }\frac{q^2}{3} v_{q}^2{1\over
2}\int_{0}^\infty \frac{d\omega '}{\pi}\frac{\chi''_{0\uparrow}(q,\omega
')  \chi''_{0\downarrow}(q,-\omega ')}{|\epsilon(q,\omega ')|^2
\sinh^2(\beta\omega '/2)}.  \end{equation}

Furthermore, for $T=0$ and $\omega\ne 0$, the RPA is equivalent to the
decoupling approximation for the four-point response function used in
\cite{Nifosi} to calculate the dynamical exchange-correlation kernel. In
this limit, the real part of the spin transresistivity takes the form
\begin{eqnarray} &~& Re \rho_{\uparrow\downarrow}(\omega,0) =
\frac{1}{n_\uparrow n_\downarrow e^2\omega} {1\over V}\sum_{\vec q
}\frac{q^2}{3} v_{q}^2\nonumber \\&~& \int_{0}^\omega \frac{d\omega
'}{\pi}[\chi''_{\uparrow\uparrow}(q,\omega ')
\chi''_{\downarrow\downarrow}(-q,\omega-\omega ')  -
\chi''_{\uparrow\downarrow}(q,\omega ')
\chi''_{\downarrow\uparrow}(-q,\omega-\omega ')] \label{rhoT=0}
\end{eqnarray} Thus our calculation demonstrates that those two
approximations, quite different at a first sight, are simply RPAs
performed in different limits.

From now on, for simplicity of notation, we will refer to $ Re
\rho_{\uparrow\downarrow}$ simply as $\rho_{\uparrow\downarrow}$.

\subsection{Numerical evaluation} \label{allT} To calculate $\rho_{\uparrow
\downarrow}$ at finite temperature, we have used in Eq.~(\ref{rho})  the
temperature dependent expression for the three-dimensional noninteracting
spin-resolved density-density response function \begin{eqnarray}
\chi_{0\alpha}''(q,\omega;T) & = &-{1 \over 16\pi}{1 \over \bar{q} a^{*3}
Ry} \cdot \nonumber \\ &~& \cdot\left\{ \bar{\omega} -{1\over \beta
Ry}\ln{1+e^{\beta\left[{1 \over \epsilon_q}\left({\hbar\omega + \epsilon_q
\over 2}\right)^2-\xi_\alpha\right]} \over 1+e^{\beta\left[{1 \over
\epsilon_q}\left({\hbar\omega - \epsilon_q \over
2}\right)^2-\xi_\alpha\right]}}\right\},\label{chi_T} \end{eqnarray} where
$a^*$ is the effective Bohr radius, $\bar{q}=qa^*$, $\bar{\omega}=\hbar
\omega/Ry$, $Ry=e^2/2a^*$ is the effective Rydberg, $\xi_\alpha$
 is the chemical potential for the $\alpha$ spin population and
  $\epsilon_q=\hbar^2q^2/2m^*$.  Eq.~(\ref{chi_T}) follows directly from
the definition \begin{equation}\label{def_chi}
\chi_{0\alpha}''(q,\omega;T)=-{\pi\over V}\sum_{{\vec k}}(n_{{\vec
k}\alpha}- n_{{\vec k}+{\vec
q}\alpha})\delta(\hbar\omega+\epsilon_{k}-\epsilon_{ k+q}), \end{equation}
where $n_{{\vec q}\alpha}=1/[\exp(\beta(\epsilon_{q}-\xi_\alpha))+1]$ is
the average number of $\alpha$-spin electrons with energy $\epsilon_{q}$.

Fig.\ref{fig2} shows $|\rho_{\uparrow \downarrow}|$ as a function of
temperature and density \cite{rhoneg}.  The data are calculated in the paramagnetic
phase and for semiconductor parameters (GaAs), i.e. $m^*=0.067$,
$\epsilon=12$ and carrier density $n_1=1.5\times 10^{16}~cm^{-3}$,
$n_2=1.5\times 10^{17}~cm^{-3}$, and $n_3=1.5\times 10^{18}~cm^{-3}$.
$\rho_{\uparrow \downarrow}$ peaks at about the Fermi temperature $T_F$,
underlying the crossing between the degenerate and the non degenerate
regimes. As can be seen, $\rho_{\uparrow \downarrow}$ is strongly enhanced
as the density decreases, mainly due to the prefactor dependence $\sim
1/n^2$. In fact its maximum increases of almost two orders of magnitude,
passing from $0.3$ milli$\Omega \times cm$ for $n=n_1$ to $14$
milli$\Omega \times cm$ for $n=n_3$. 
In the calculations of the following sections, we will
mainly focus on the density value $n=n_2$, corresponding to a Fermi
temperature $T_F=178K$.  The inset of Fig.~\ref{fig2} presents for this
density value
 the comparison between $\rho_{\uparrow \downarrow}$ and its
non-degenerate analytical approximation (dashed lines) discussed in
details
 in Sec.\ref{ndeg}.

We now turn to a quantitative assessment of the relevancy of the spin
Coulomb drag.  First of all it is necessary to underline that the spin drag
is an {\it intrinsic} effect of  spin-polarized transport: that is,  while  impurity
scattering could in principle be suppressed in a perfect crystal, the spin
Coulomb drag will always be present, even in the purest sample, and 
dominate over phonon scattering at sufficiently low temperature.  However,
since available samples are usually far from perfection, it is reasonable to ask
how the spin transresistivity compares to the more familiar Drude resistivity.  In
metals, as we shall show in detail
 in the next section,  one finds, at most, $\rho_{\uparrow \downarrow}\sim
10^{-2}\mu\Omega\times cm$
 so that $\rho_{\uparrow \downarrow}/\rho_D$ is of the order of few
percents.  The situation is very
 different for semiconductors: since both the Fermi temperature (at which
$\rho_{\uparrow \downarrow}$ peaks) and the carrier density are
considerably lower than in metals, $\rho_{\uparrow \downarrow}$ can become
comparable and even greater than $\rho_D$. This strong variation depends
on  specific semiconductor characteristics,  such as the effective mass, 
mobility, and density  of the carriers.

In Fig.~\ref{fig3} we show the effect of the carrier mobility $\mu$ on the ratio
$\rho_{\uparrow \downarrow}/\rho_D$. We plot $\rho_{\uparrow
\downarrow}/\rho_D$ in respect to temperature for an n-doped semiconductor
as GaAs ($m^*=0.067m_e$, $\epsilon=12$ and $n=1.5\times
10^{17}~cm^{-3}$; upper panel) and for a p-doped semiconductor, as
(Ga,Mn)As
 ($m^*=0.5m_e$, $\epsilon=12$, $n=1.2\times 10^{19} cm^{-3}$; lower
panel). Each curve corresponds to a different mobility value as reported
in the figure caption. The values increase from "A"  to "D".  In
particular the value $\mu=3\times 10^3 cm^2/Vs$ (labels as "C"),
corresponds to the value measured for a spin packet in \cite{Awscha5}.
 As can be seen, changing the material it is possible to increase the
ratio $\rho_{\uparrow \downarrow}/\rho_D$ by an order of magnitude, to the
point that the spin transresistivity can become greater than the Drude
resistivity.

In Fig.~\ref{fig4} we show the dependence on the ratio $\rho_{\uparrow
\downarrow}/\rho_D$  on the carrier density for GaAs. The results are
presented for two different temperatures ($T=20K$, dashed lines and
$T=300K$, solid line) and two different values of the mobility (as
labelled in the figure). Each curve peaks at a 
 density such that $T_F = T$.

In conclusion,  our calculations demonstrate that 
 in semiconductors, at $T\approx T_F$,  the spin Coulomb
drag must definitely be considered an important contribution to the 
resistivity for spin polarized currents.

\subsection {The degenerate limit\label{deg}} Let us now focus on the
low-temperature ($k_BT << E_F$) and low-frequency ($\hbar\omega <<E_F$)
regime.  This is the only regime of practical importance in ordinary high-
density metals.  The spin-Coulomb drag coefficient is controlled by a
subset of all the processes that lead to the finite lifetime of a
quasi-particle at the Fermi surface, namely, the processes in which the
quasi-particle in question exchanges momentum with an electron of opposite
spin  (scattering processes between
parallel spin electrons do not cause relaxation of the spin-current).
Since the inverse quasi-particle life-time at the Fermi surface is known
in Fermi liquid theory to scale as $(k_BT/E_F)^2$ we expect the same
scaling to hold for the spin drag coefficient at low temperature.

This prediction is confirmed by the detailed calculation as follows.  At
low temperature, the exponential factors in Eq.~(\ref{rho}) restrict the
region of integration to $\omega \sim k_BT/\hbar$.  The low-frequency form
of the density fluctuation spectra $\chi''_{\alpha \alpha'}(q,\omega)$ is
a linear function of $\omega$.  In the limit of vanishing impurity
concentration $\chi_{0 \alpha}(q, \omega)$ is simply the Lindhard
function, whose imaginary part, at low frequency, is given by
$\chi_{0\alpha}''({\vec q},\omega\to 0)=-(m^{*2}/4\pi)(\omega/q)$ and whose
real part can be approximated by its value at $\omega=0$.  Making use of
these limiting forms, the calculation of $\rho_{\uparrow \downarrow}$ can
be carried in an essentially analytical fashion. The
  result is 
\begin{eqnarray} Re\rho_{\uparrow\downarrow}(\omega,T)
&=&-{\hbar a^* \over e^2}{4\pi^2 (k_BT)^2 + \hbar^2\omega^2 \over 6
(Ry)^2}\cdot \nonumber \\ &~&{1\over 24\pi^3
\bar{n}_\downarrow\bar{n}_\uparrow} \int_{0}^{2k_Fa^*}{d\bar{q}\over
\bar{q}^2} {1\over |\epsilon (\bar{q}/a^*,0)|^2} \label{rholim} ~,
\end{eqnarray} 
where $k_F\equiv \min (k_{F\uparrow},k_{F\downarrow})$,
with $k_{F\alpha}$ the $\alpha$ spin population Fermi wave-vector,
 $\bar{n}_\alpha\equiv n_\alpha a^{*3}$ and
 $\epsilon (q,\omega) = 1-v_q\chi_{0\uparrow}({\vec q},
\omega)-v_q\chi_{0\downarrow} ({\vec q}, \omega)$ is the RPA dielectric
function.  Eq.~(\ref{rholim}) shows that, in the absence of impurities,
$\rho_{\uparrow\downarrow}(\omega,T)$ is proportional to $\omega^2$ for
$k_BT\ll \hbar\omega$ and to $T^2$ for $\hbar\omega\ll k_BT$.

Modifications in the form of $\chi_{0 \alpha}(q, \omega)$ due to the
presence of impurities can be taken into account through Mermin's
approximation scheme \cite{Mermin}. These modifications amount to
replacing $\omega/q v_F$ by $\omega /Dq^2$ ($D = v_F^2 \tau/3$ being the
diffusion constant) for $\omega < 1 /\tau$ and $q<1/v_F \tau$ where $v_F$
is the Fermi velocity and $\tau$ is the electron-impurity mean scattering
time.  The $\omega$ and $T$ dependencies of Eq.~(\ref{rholim})  are not
affected.

Writing explicitly in Eq.~(\ref{rholim}) the dependence on $r_{s\alpha}$
(where $r_{s\alpha}=(4\pi n_\alpha a^{*3}/3)^{-1/3}$ is the value of $r_s$
for spin $\alpha$) one can also see that
$\rho_{\uparrow\downarrow}(\omega,T) \sim
r_{s\uparrow}^3r_{s\downarrow}^3$, so that $ |\rho_{\uparrow\downarrow}| $
will strongly increase with decreasing electron density. In Fig. \ref{fig5} we
plot $ |\rho_{\uparrow\downarrow}(\omega=0,T)|$ as a function of the
temperature, for $n_\uparrow=n_\downarrow$ and in the density range
$1<r_s<7$. The Figure shows
 that, for metallic densities correspondent to $r_s\stackrel{>}{
 \sim}5$ and temperatures of the order of $40-60K$ (at which for example
experiments on spin relaxation time using spin polarized currents have
been performed \cite{Silsbee}), the spin trans-resistivity is appreciable
($ |\rho_{\uparrow\downarrow}(\omega=0,T)|\stackrel{>}{
 \sim} 0.01\mu\Omega cm$).  The corresponding Coulomb scattering time is
$\gamma^{-1}$ from Eq.~(\ref{gamma}).  For $r_s=5$ we obtain
$\gamma^{-1}\approx 10^{-13}s$ and $\delta_s/v_F\approx 10^{-10}s$:
$\gamma^{-1}$ is indeed several orders of magnitude smaller than the
spin-flip time. This demonstrates that neglecting spin-flip processes is
indeed a good approximation for this kind of metals.

\subsection {The non-degenerate limit} \label{ndeg} 
The non-degenerate
limit is characterized by $T>>T_F$.  First of all we calculate
 the non-degenerate limit of the non-interacting temperature-dependent
spectral function Eq.~(\ref{chi_T}): starting from the definition
Eq.~(\ref{def_chi}), we use the classical expression for the fugacity
$\exp(\beta\xi_\alpha)  =n_\alpha\cdot 8\pi^3(\beta/2m^*\pi)^{3/2}$, and
 obtain \begin{equation}\label{chiTnd}
\chi_{0\alpha}''(q,\omega;T)=-{\sqrt{2\pi\beta m^*}n_\alpha \over \hbar
q}\exp(-{\beta\epsilon_q \over 4})  \exp(-{\beta \hbar^2\omega^2 \over 4
\epsilon_q})\sinh({\beta\hbar\omega \over 2}).  \end{equation} In
addition, in order to calculate the non-degenerate limit $\rho_{\uparrow
\downarrow}$, we have used
  the classical limit for the dielectric constant $\epsilon (q,\omega)= 1+
(4\pi e^2/\epsilon q^2)(n/k_BT)$.  The final result is
\begin{eqnarray}\label{nondegrho}
 \rho_{\uparrow \downarrow}(0,T)&=&{8e^2\sqrt{m^*} \over \sqrt{2\pi^3}
(k_BT)^{3\over 2} \epsilon^2} \int_0^\infty dx{x\exp(-x) \over
(x+\lambda)^2} \\ &\approx & {8e^2\sqrt{m^*} \over \sqrt{2\pi^3}
(k_BT)^{3\over 2} \epsilon^2} [-1-C-\ln(\lambda)],\label{nondegrho2}
\end{eqnarray} where the second expression, Eq.~(\ref{nondegrho2}),
  is valid in the limit $\lambda\ll 1$, $\lambda =\hbar^2 k_D^2/k_BT4m^*$,
$k_D^2\equiv 4\pi e^2 n/\epsilon k_BT $ is the inverse of the squared Debye
screening length and $C\approx 0.577$ is the Euler's constant.  Notice
that, in the non-degenerate limit, $\rho_{\uparrow\downarrow}$ becomes
almost independent of the total density $n$ and independent of the spin
density
 components $n_\alpha$, while a quantum mechanical dependence on $\hbar$
survives even in this regime. $\rho_{\uparrow\downarrow}$ tends to zero as
$(k_BT)^{-3/2}\ln(k_BT)$
 as $T\to\infty$.  The inset of Fig.~\ref{fig2} illustrates the comparison
between $ \rho_{\uparrow \downarrow}$ and its asymptotic form. This
approximation becomes valid
 for $T\gg T_F$, but, since $T_F\sim n^{2/3}$,
 such limit is fulfilled  only at very low
carrier densities.

\subsection {The mixed (degenerate/nondegenerate) case}

 A very interesting limit is the one corresponding to
 a spin polarization process,
 for which $n_\uparrow \to n$ and $n_\downarrow \to 0$. This is indeed
 relevant for one of the problems
 we want to analyze, i.e. a semiconductor with strongly spin-polarized carriers. 
This system is in a very peculiar state: its minority down-spin
population is non-degenerate, i.e. $k_BT\gg E_{F\downarrow}$,
 $E_{F\alpha}=\hbar^2(6\pi^2n_{\alpha})^{2/3}/2m^*$, while, for low enough
temperatures, the majority up-spin population is degenerate, i.e. $k_BT\ll
E_{F\uparrow}$. The expression for the non-interacting spin-resolved
density-density response
 functions entering the spin transresistivity Eq.~(\ref{rho}) can then be
taken from the previous sub-sections and are given by
\begin{eqnarray} \chi_{0\uparrow}''(q,\omega';T)&=&-{m^{*2}\over 4\pi
\hbar^3}{\omega'\over q} \mbox{~~~~~for
$0<q<2k_{F\uparrow}$}\label{chiu2}\\ &=& 0
\mbox{~~~~~~otherwise}\label{chiu1} \end{eqnarray} and
\begin{equation}\label{chid} \chi_{0\downarrow}''(q,\omega';T)=-\left({\pi
m^*\over 2}\right)^{1\over 2}\beta^{3\over 2} n_\downarrow\exp(-{\beta
m^*\omega'^2\over2q^2}){\omega'\over q}, \end{equation} where
Eq.~(\ref{chiu1}) is valid up to first order in $\omega'$ and
Eq.~(\ref{chid})
 represents the classical limit ($\hbar\to 0$) of Eq.~(\ref{chiTnd}).
Using Eqs.~(\ref{chiu1}), (\ref{chid}) and the approximation, due to the
small density of down-spin carriers
$\epsilon(q,\omega')\approx\epsilon_\uparrow(q,0)$,
 the expression for the spin transresistivity becomes \begin{equation}
Re\rho_{\uparrow\downarrow}(0,T) =-{\hbar a^* \over e^2}{2\sqrt{\pi} \over
9} {(k_BT)^{1\over 2}\over(Ry)^{1\over 2}} {1\over \bar{n}_\uparrow}
\int_{0}^{2k_{F\uparrow}a^*}{d\bar{q}\over \bar{q}^2} {1\over
 |\epsilon_\uparrow (\bar{q}/a^*,0)|^2}.  \label{rhomix} \end{equation}
Eq.~(\ref{rhomix}) is very similar to Eq.~(\ref{rholim}),
 the result obtained when both components are degenerate. This is a
consequence of the fact that, in the appropriate regime, the
non-degenerate spectral function Eq.~(\ref{chid}) presents the same
dependence in $\omega'$
 and $q$ as the degenerate one. 
We want to underline that, in this limit, $\rho_{\uparrow\downarrow}(0,T)$
is independent of $n_\downarrow$ and that in any case
$\rho_{\uparrow\downarrow}\to 0$ for $T\to 0$.

\section {Spin stiffness of an interacting spin-polarized electron gas}
The other ingredient entering the drift-diffusion expression for the
current, Eq.~(\ref{driftdiffusion}), is  the longitudinal spin stiffness matrix
$S_{\alpha\beta}$. In particular we are interested in the combination $S =
\partial^2 f(n,m,T)/\partial m^2=(S_{\uparrow\uparrow}-
S_{\uparrow\downarrow}+S_{\downarrow\downarrow}- S_{\downarrow\uparrow})/4
$,  where $m=n_\uparrow-n_\downarrow$, 
which gives the curvature of the free energy with respect to the
magnetization  at constant density.
This quantity coincides with the inverse of the
longitudinal spin susceptibility of the uniform electron gas.

We evaluated  $S$ numerically starting from the formulas provided by
Tanaka and Ichimaru \cite{Ichimaru}, who calculated the free energy density
of the three-dimensional electron gas as a function of temperature,
density, and
spin polarization.

Fig.\ref{fig6} shows $S$ divided by its non-interacting value $S_{ni}$ as a
function of the dimensionless temperature  $T/T_F$ for various densities,
starting with  $n=4.2 \times 10^{17} cm^{-3}$ for  the upper curve down to
$n=4.2 \times 10^{11} cm^{-3}$ for the lowest one, decreasing by an order
of magnitude from one curve to the next.

Two regimes are clearly visible.   For densities larger than  a critical
value $n_c  \simeq 4.2 \times 10^{13} cm^{-3} $  the spin
stiffness decreases monotonically with decreasing temperature settling to a
finite value in the ground-state (top four curves).
  For densities lower than $n_c$ a
second-order ferromagnetic transition occurs:  the critical
temperature  $T_c$ raises from $\sim 0$ at $n=n_c$ to a sizeable fraction
of the
Fermi temperature at  $n=4.2 \times 10^{11} cm^{-3}$.  As in any
second-order transition, the spin stiffness vanishes at the transition
temperature \cite{history}.  
 For $T<T_c$ the spontaneous  magnetization $\bar m$ is given by the {\it
stable }  minimum of the free energy, which satisfies the conditions
\begin{eqnarray}
&& {\partial f(n,m,T) \over \partial m} \biggl  \vert_{m=\bar m}  = 0,
\nonumber \\
&& \bar S = {\partial^2 f(n,m,T) \over \partial m^2} \biggl \vert_{m=\bar
m} >0.  \end {eqnarray}
Obviously, ferromagnetism shows up only at extremely low densities, and the
electron gas model may  break down well before getting to such densities:
it is
nevertheless instructive to study the repercussions of the behavior of $S$
on the
spin diffusion constant both above and below the critical density.

Let us now consider the limiting behaviors of the spin stiffness at high
($T>>T_F$) and low ($T<<T_F$) temperatures.

\subsection {The high temperature limit} \label{nondegS} In the
high-temperature limit ($T\gg T_F(n)$) the free energy density of the electron
gas has the following expansion:
\begin{eqnarray} \label{hightfreeenergy}
f(n_\uparrow, n_\downarrow, T)  &\simeq&   k_BT \sum_\alpha n_\alpha [\ln
(n_\alpha \lambda_T^3) -1] \nonumber \\
&+& {k_BT \over 2^{3/2}} \sum_\alpha n^2_\alpha  \lambda_T^3
\nonumber \\
&-& 2\pi^2 e^2 \lambda_T^2 \sum_\alpha n_{\alpha}^2
\nonumber \\ 
&-& {2 \pi^{1/2} \over 3}{e^3 n^{3/2} \over (k_BT)^{1/2}},
\end{eqnarray}
where $\lambda_T \equiv (2 \pi \hbar^2/m^*k_BT)^{1/2}$ is the thermal
wavelength. The first term is the free energy of the classical ideal gas, the
second term is the leading quantum correction for noninteracting Fermions
\cite{huang}; the third term is the leading quantum/interaction
correction, namely, the high temperature exchange free-energy\cite{LL},  the last
term is
the leading classical interaction correction from Debye-Huckel theory.  Only
the first three terms depend on the magnetization, and therefore contribute
to the
spin stiffness. Taking a second derivative with respect to magnetization we
find,  after simple calculations,
 \begin{equation} \label{hightstiffness}
{S \over S_c} = 1 + {n_\uparrow n_\downarrow \over n^2}  \biggl [
{n \lambda_T^3 \over 2^{1/2}} - {8\pi^2 e^2 n^{1/3} \over k_B T}
(n \lambda_T^3)^{2/3} \biggr ] ,
\end{equation}
where  $S_c$ is the Curie spin stiffness of an ideal classical gas of density
$n$:
\begin {equation} S_c = { k_BT n \over 4
n_\uparrow n_\downarrow}.\label{SCurie} \end{equation}
Notice that the leading interaction correction to the noninteracting spin
stiffness is
negative, in agreement with the behavior seen in Fig. 6.

\subsection {The low temperature limit } \label{mixlimS}
Let us now examine what happens in the limit  $T \to 0$.  Above the
critical density
$n_c$ the spin stiffness simply tends to a constant zero-temperature limit,
smaller
than the noninteracting value,    in agreement with the Landau theory of Fermi
liquids.

For $n<n_c$ the low temperature phase is ferromagnetic, and in this case the
density of majority spin electrons ($n_\uparrow$) approaches the total
density, while
the density of minority spin electrons   ($n_\downarrow$) tends to zero for $T
\to 0$.   To understand the behavior of the spin stiffness shown in Fig. 6 we
assume that, in the nearly $100 \%$ polarized limit the free energy can be
written as the sum of the ground-state energy of the degenerate interacting
up-spin
gas plus the free energy of  an infinitely dilute noninteracting down spin gas:
\begin{equation} \label{zerof}
f(n_\uparrow, n_\downarrow, T) \simeq \epsilon_0(n_\uparrow)
- k_BT n_\downarrow \ln (1 - \zeta) + constant,
\end{equation}
where $\epsilon_0(n)$ is the ground-state energy density of a  degenerate
liquid of up-spin electrons, and  $\zeta =( n_\uparrow - n_\downarrow)/(
n_\uparrow + n_\downarrow)$ is the degree of spin polarization.  Obviously,
this
approximation  ignores the correlation between down- and up-spin electrons, or,
more precisely, presumes that this correlation is smaller than the entropic
term
$k_BT n_\downarrow \ln (1 - \zeta) $ for $\zeta \to 1$.

Starting from  Eq.~(\ref{zerof}) it is trivial to show that the  minority
spin density
vanishes for $T \to 0$ as
\begin{equation}
n_\downarrow \sim n  e^{\epsilon_0'(n)/k_BT}
\end{equation}
where $\epsilon_0'(n) = d\epsilon_0(n)/dn  <0$ at low density, while the
spin stiffness goes as
\begin {equation}
S \sim {k_BT \over 4 n_\downarrow},
\end{equation}
 which diverges exponentially for $T \to 0$.  This is precisely what our
numerical
calculations, based on the formulas of Ref.~(\cite{Ichimaru}), indicate.
As we shall see, this result is important in understanding
the behavior of the diffusion constant of a unipolar spin packet when the
system is
fully spin polarized.

\section {The evolution of a spin packet} \label{secevo} 
We  now
examine  the motion of a spin
packet under the effect of a uniform electric field.  Let us apply
Eq.~(\ref{driftdiffusion}) to calculate the time evolution of a spin
packet obtained by injecting an {\it excess} spin density $\Delta m(\vec
r, 0) = \Delta M \delta (\vec r)$ near the origin at time $t=0$. We denote
by $ m (\vec r,t) = n_{\uparrow} (\vec r,t) - n_{\downarrow} (\vec r,t)$
the net spin density at point $\vec r$ and time $t$, by $m^{(0)} =
n_{\uparrow}^{(0)} - n_{\downarrow}^{(0)}$ the uniform value of the spin
density at thermodynamic equilibrium, and by $\Delta m( \vec r ,t) \equiv
m(\vec r,t) -m^{(0)}$ the excess spin density following spin injection.
To solve this problem we combine the expression for the current density
Eq.~(\ref{driftdiffusion}) with the generalized continuity equations for
the spin-density components
\begin{equation} \label{spincontinuity}
{\partial \Delta n_\alpha(\vec r,t) \over \partial t} = - {\Delta n_\alpha
(\vec r,t) \over \tau_{sf, \alpha}} + {\Delta n_{\bar{\alpha}}
(\vec r,t) \over \tau_{ sf, \bar \alpha}} - \vec \nabla \cdot \vec
J_\alpha(\vec r), \end{equation} where $\tau_{sf, \alpha}$ is the
spin-flip relaxation time for the $\alpha$ component.  Substituting in
this equation the drift-diffusion expression for the current
Eq.~(\ref{driftdiffusion})  we obtain the two equations
\begin{eqnarray} {\partial \Delta
n_\alpha(\vec r,t) \over \partial t} &=& - {\Delta n_\alpha (\vec r,t)
\over \tau_{sf, \alpha}} + {\Delta n_{\bar{\alpha}} (\vec r,t)
\over \tau_{sf, \bar \alpha}}\nonumber\\& & +{E\over e}\sum_\beta
{\partial \tilde{\sigma}_\alpha \over \partial n_\beta}\nabla(\Delta
n_\beta)+\tilde{\sigma}_\alpha{\nabla\cdot E\over e}
+\sum_\beta\left[\nabla D_{\alpha\beta}\nabla
n_\beta+D_{\alpha\beta}\nabla^2(\Delta n_\beta)\right]\label{ddalpha},
\end{eqnarray} 
($\alpha = \uparrow$ or $\downarrow$) where 
\begin{equation}\label{sig_til}
\tilde \sigma_{\alpha} =\sum_\beta \sigma_{\alpha\beta}.  \end{equation}
Eq.~(\ref{ddalpha})  includes the term $\tilde{\sigma}_\alpha\nabla\cdot E/
e= \tilde{\sigma}_\alpha(\Delta n_\uparrow+\Delta n_\downarrow)/\epsilon$.
This term is ``dangerous'' because it contains the product of a large
quantity $\tilde{\sigma}_\alpha$ times a small quantity, the space charge
$\Delta n_\uparrow+\Delta n_\downarrow$,  the product itself being of the
order of the other quantities of interest in the calculation.

It is tempting, but wrong, to invoke the local charge neutrality
constraint \begin{equation} \label{chargeneutrality} \Delta n_{\uparrow}
(r) = - \Delta n_{\downarrow} (r)  \end{equation}
 at this point.   Instead, we will first combine the two
component of Eqs.~(\ref{ddalpha}) to eliminate the space charge ($\nabla\cdot
E$) term, and only after doing that can we impose, without  serious loss of
accuracy,
the charge neutrality constraint. To eliminate the $\nabla \cdot E$ term we
multiply
each component of Eq. (\ref{ddalpha}) by the conductivity $\tilde
\sigma_{\bar{\alpha}}$ of the opposite channel, and, in order to get the
equation of
motion for $\Delta m$, we subtract the equation for $\alpha$-spins
 from the  equation for $\bar \alpha$-spins.  Only at this point
we impose the local charge neutrality constraint
Eq.~(\ref{chargeneutrality}).  With this procedure we obtain the correct
drift-diffusion equation for $\Delta m( \vec r ,t)$ \begin{eqnarray}
{\partial \Delta m(\vec r,t) \over \partial t} & = & - {\Delta m (\vec
r,t) \over \tau_s} + {\sum_\alpha \tilde \sigma_{\bar{\alpha}} \nabla
(\tilde D_{\alpha}\nabla\Delta m(\vec r,t) )\over \sum_\alpha \tilde
\sigma_{\alpha}}\nonumber \\ & + & {\sum_\alpha \tilde
\sigma_{\bar{\alpha}} \tilde \mu_{\alpha} \over \sum_\alpha \tilde
\sigma_{\alpha}} \vec E \cdot \vec \nabla \Delta m(\vec
r,t)\label{packeteomgen} \end{eqnarray} where $\tau_s=(1/\tau_{sf, \uparrow
} +1/\tau_{sf, \downarrow})^{-1}$ is the spin relaxation
time, which is very long \cite{Awscha5,Silsbee}, \begin{eqnarray}
 \label{Dtilde} \tilde D_\alpha &=& D_{\alpha \alpha}- D_{\alpha \bar
\alpha},\\ \label{mutilde} \tilde \mu_\alpha &=& \mu_{\alpha
\alpha}+\mu_{\alpha \bar \alpha}, \end{eqnarray} $\vec E$ is an externally
applied electric field, and the matrix $\mu_{\alpha \beta}$ is defined as
\begin{eqnarray} \label{mobilitymatrix} e \mu_{\alpha \beta} & \equiv &
{\partial \sigma_{\alpha \beta} \over \partial n_{ \alpha}}-{\partial
\sigma_{\alpha \beta} \over \partial n_{ \bar{\alpha}}} \\ & = &\pm 2
{\partial \sigma_{\alpha \beta} \over \partial m},~ \mbox{plus if
$\alpha=\uparrow$, minus otherwise.} \end{eqnarray} This is a
generalization of the familiar relation between mobility and conductivity:
it takes into account the dependence of the mobility $\mu_{\alpha \beta}$
 on both spin density components. The second term in
Eq.~(\ref{mobilitymatrix}) accounts for
 the reduction of the mobility in the $\alpha$ channel due to the drag of
the $\alpha$ spin population on the $\bar{\alpha}$ population
\cite{spindrag}.

The fact that the mobilities enter Eq.~(\ref{packeteomgen}) as
 a spin symmetric combination (Eq.~(\ref{mutilde})) while the diffusion
constants are in a spin antisymmetric combination (Eq.~(\ref{Dtilde})),
 reflects the fact that the electrostatic field has the same sign for both
spin components, while the density gradients have opposite signs (see
Eq.~(\ref{chargeneutrality})).

If we consider the linear regime - i.e. we neglect terms of the order of
$(\nabla n_\alpha)^2$ -
  Eq.~(\ref{mobilitymatrix}) reduces to the more familiar
\begin{equation}\label{sigmamu} \sigma_{\alpha \beta}= e
n_\alpha\mu_{\alpha \beta} \end{equation} and Eq.~(\ref{packeteomgen}) can
be written as \begin{equation} \label{packeteom} {\partial \Delta m(\vec
r,t) \over \partial t} = - {\Delta m (\vec r,t) \over \tau_s} + D_s
\nabla^2 \Delta m(\vec r,t) + \mu_s \vec E \cdot \vec \nabla \Delta m(\vec
r,t), \end{equation} where \begin{equation} \label{muspin}
 \mu_s = {\sum_\alpha \tilde \sigma_{\bar{\alpha}} \tilde \mu_{\alpha}
\over \sum_\alpha \tilde \sigma_{\alpha}} \end{equation} and
\begin{equation} \label{Dspin}
 D_s = {\sum_\alpha \tilde \sigma_{\bar{\alpha}} \tilde D_{\alpha} \over
\sum_\alpha \tilde \sigma_{\alpha}}, \end{equation} are the effective
mobility and diffusion constants \cite{footnote3}.  We underline that in
reality the range of validity of Eq.~(\ref{packeteom})  extends beyond the
linear approximation into the classical regime, i.e.
 to high carrier densities or high temperatures, since in that regime the
relationship between density and conductivity is linear.
Eqs~(\ref{muspin}) and (\ref{Dspin}) show that the mobility and the
diffusion constants of the packet are weighted averages of, respectively,
the mobilities $\tilde \mu_{\alpha}$ and diffusion constants $\tilde
D_{\alpha}$ of the two spin channels, the weight being the conductivity
$\tilde \sigma_{\bar{\alpha}}$
 of the {\it opposite} channel.  This is due to the local charge
neutrality constraint Eq.~(\ref{chargeneutrality}), that forces the two
components of the disturbance to travel together, so that the conductivity
of each spin channel is strongly influenced by the motion of the disturbance
in the other channel.  In the non interacting limit Eqs~(\ref{muspin}) and
(\ref{Dspin}) reduce to the expressions presented in \cite{Flatte}.

The solution of Eq.~(\ref{packeteom}) is \begin{equation}
\label{packetevolution} \Delta m (\vec r,t) = {\Delta M e^{-t/\tau_s}
\over (4 \pi D_s t)^{3/2}} e^{-{|\vec r + \mu_s \vec E t|^2 \over 4 D_s
t}}.  \end{equation} Eq.~~(\ref{packetevolution}) has the form of a
Gaussian packet that drifts
 under the effect of the electric field $\vec{ E}$ with a pace determined
by $\mu_s$, and spreads in time at a rate determined
 by $D_s$.  The mobility and diffusion constants of electron-hole packets
of similar shape can be measured through the Haynes-Shockley experiment
\cite{Shockley}.  Thus a similar experiment can in principle determine
 $\mu_s$ and $D_s$ independently, provided that $\tau_s$ is sufficiently
long.

With the help of some algebra, Eq.~(\ref{muspin})  can be rewritten as
\begin{equation}\label{muspin2} \mu_s = {1\over e}{n\over n_\uparrow
n_\downarrow}{1\over\sum_\alpha(1/ \tilde\sigma_\alpha)} \end{equation}
where $\tilde{\sigma}_\alpha$ is given by Eq.~(\ref{sig_til}),
 and Eq.~(\ref{Dspin}) as \begin{eqnarray}\label{Dspin2}
 D_s & = & {S\over e^2} {4\over \sum_\alpha \tilde \rho_{\alpha}}\\ & = &
{k_BT\over e^2n}\cdot{S\over S_c}\cdot {1\over [(n_\uparrow n_\downarrow/
n^2)(\rho_{D\uparrow}
+\rho_{D\downarrow})-\rho_{\uparrow\downarrow}]},\label{Dspin3}
\end{eqnarray} where \begin{equation} \tilde \rho_{\alpha}\equiv
\rho_{\alpha\alpha}-\rho_{\alpha\bar{\alpha}}, \end{equation} $S_c$ is
given by Eq.~(\ref{SCurie}), $\rho_{D\alpha} =m^*/ ne^2 \tau_{\alpha}$ is
the ordinary Drude resistivity associated with the $\alpha$ spin channel,
and $\rho_{\uparrow \downarrow}$ - a negative number - is the spin drag
transresistivity discussed in Sec. \ref{secsd}.

Eq.~(\ref{muspin2})  can be seen as the generalization to a spin packet of
the ordinary relationship between mobility and conductivity, while
Eq.~(\ref{Dspin2}) is the corresponding
 generalization of the Einstein relation.  To derive Eq.~(\ref{Dspin3}),
we made use explicitly of the structure of the matrix $\rho_{\alpha\beta}$
discussed in Sec. \ref{secstrusd}.  

The calculation of $\mu_s$ and $D_s$ is simplified under the
assumption that the scattering times for the two spin components are not
too different, i.e., $\tau_{\uparrow} = \tau_{\downarrow} = \tau_D$. This
assumption is well justified for non-degenerate carriers.  In the
degenerate case $1/\tau_\alpha$ has a weak $n_\alpha^{1/3}$ dependence on
the density.  From Eqs.~ (\ref{muspin}) and
(\ref{Dspin3})  we obtain \begin{equation} \label{musfinal} \mu_s = {e
\tau_D \over m^*} \end{equation} and \begin{equation} \label{dsfinal} D_s
= {\mu_s k_BT \over e} {S \over S_c} {1 \over 1 - \rho_{\uparrow
\downarrow}/\rho_D}, \end{equation} where $\rho_D =m^*/ ne^2 \tau_D$ is
now the ordinary Drude resistivity.

Eq.~(\ref{musfinal}) tells us that the mobility of the packet is not {\it
explicitly} modified by Coulomb interaction and in fact coincides with the
ordinary homogeneous mobility.  Strictly speaking this result is only
valid under the assumption that up-spin and down-spin electrons have equal
mobilities and thus drift at the same speed in an applied electric field.
Coulomb interactions, being Galilean-invariant, cannot change the total
momentum of such a uniformly drifting electron gas.

The situation is completely different for the diffusion constant. As the
spin packet spreads out the up- and down-spin currents are directed in
opposite directions and friction arises: for this reason the expression
for $D_s$ contains the spin-drag resistivity as a factor that reduces the
diffusion.
  In addition, the Coulomb interaction together with the Pauli exclusion
principle reduces the energy cost of spin-density fluctuations (i.e., the
spin stiffness) decreasing further the rate of diffusion of a spin packet.

\section {Experimental observation of interaction effects in
spin-polarized transport} \subsection {Direct observation of spin Coulomb
drag} \label{secsdexe} We will now propose an experiment aimed at
detecting the effect of the spin Coulomb drag and measuring the spin
trans-resistivity.  We describe the experiment as it could be done on
metals, but the same scheme
 could be applied to semiconductors (in which the drag effect is larger),
 provided that an appropriate method of injecting spin current is used.

The setup is shown in Fig.~\ref{fig7}: a paramagnetic metal film of
thickness $L$ is sandwiched between two ferromagnets polarized in the same
 direction. A battery is connected to the ferromagnets inducing a {\it
spin-polarized} current \cite{Silsbee} from the first ferromagnet
(``injector'')  through the paramagnet and toward the second ferromagnet
(``receiver''). The injector and receiver are chosen to be {\it
half-metals}, i.e, they have only electron states of spin $\uparrow$ at
the Fermi level (see Fig.\ref{fig7}).  It follows that the injected
current ${\vec j}_\uparrow$ is carried {\it only} by spin $\uparrow$
electrons \cite {footnote4}. If we choose $L<<\delta_s$, where $ \delta_s$
is the
spin relaxation length, we can safely neglect spin-flip processes. Spin
relaxation
lengths are relatively large in some materials ($ \delta_s\approx 100\mu m$
in Al
\cite{Silsbee}), so the condition $L<<\delta_s$ is not particularly
restrictive.
At any rate, we have seen in Section II that the coupling between spin channels
induced by spin-flip scattering is expected to be  much weaker than the one
due to
the Coulomb interaction.  Due to the spin Coulomb drag, the injected ${\vec
j}_\uparrow$ will drag spin $\downarrow$ electrons toward the junction with the
receiver. But, since there is no conduction band available in the receiver
for spin
$\downarrow$ electrons the circuit will behave as an {\it open circuit} for
spin
$\downarrow$ electrons, i.e., ${\vec j}_\downarrow=0$.  The vanishing of ${\vec
j}_\downarrow$ is an indication that the Coulomb drag force is exactly
balanced by
the gradient of the electro-chemical potential for spin down electrons
\begin{eqnarray} \label{forcebalance}   \vec E_{\downarrow} &=& {\vec \nabla
\psi_\downarrow \over e} \nonumber \\ &=& \rho_{\downarrow \uparrow} \vec
j_{\uparrow} \end{eqnarray}
What Eq.~(\ref{forcebalance})
tells us is that, due to the spin Coulomb drag, there will be a
measurable electro-chemical potential difference
 $eE_{\downarrow} l = e \rho_{\downarrow \uparrow} j_\uparrow l $ for spin
$\downarrow$ electrons between two points within the metal separated by a
distance $l$ along the direction of the current.

To measure this potential difference a second circuit including a
voltmeter of very large resistance is connected to the regions of the
paramagnet close to the junctions (See Fig.~\ref{fig7}). Our purpose is to
measure $E_\downarrow$, so this second circuit must be driven by the spin
$\downarrow$ electro-chemical potential only.  In order to accomplish
this, we propose to use as contacts two half-metallic ferromagnetic
electrodes (``detectors''), similar to the injector and the receiver, but
polarized in the {\it opposite} direction. In this way, for the same
reasons explained before, the detection circuit will be ``open"  as far as
spin $\uparrow$ electrons are concerned, and the current flowing in the
voltmeter will be exclusively driven by the electro-chemical potential
difference of spin $\downarrow$ electrons.  The spin trans-resistivity
will then be given by $\rho_{\uparrow\downarrow}= (\Delta V_D
/I_\uparrow)(A/l)$, where $\Delta V_D$ is the voltage measured by the
meter, $A$ is the cross-section of the paramagnetic metal, $l$ is the
distance between the detectors, and $I_\uparrow$ the current flowing
between injector and receiver.  As shown by our calculations (see
Fig.~\ref{fig5}), we expect, in metals, a resistivity of the order of
$10^{-2}\mu\Omega cm$ that is proportional to $T^2$ for $k_B T>>\omega$.
\subsection {Haynes-Shockley experiment} The Haynes-Shockley
experiment\cite{Shockley} demonstrated the drift and diffusion of minority
carriers in a doped semiconductor.  The experiment allows a direct and
independent measure of the minority carrier diffusion and mobility coefficients. 
After a pulse of excess  carriers is created at some point in the
semiconductor, it  drifts  under the action of  an electric field, for a known
distance $L$, after which it is monitored. By measuring the drift time and the
width of the packet, it is then possible to compute both the mobility and the
diffusion constant of the packet, which coincide with those of the minority
carriers.   

The experiments  of Ref. (\cite{Awscha5})  can be seen
as a Haynes-Shockley-type experiment, since they are based on a direct
monitoring in space and time of unipolar spin packets. In these
experiments,  Kikkawa and Awschalom were able to measure
independently the diffusion and the mobility of the spin packets.  In the next
section we are going to compare our theoretical results with their
experimental findings.

\section{Paramagnetic semiconductors and spontaneous ferromagnetic
transition} \label{secpara} 
We will now focus on the results of our
calculations for the diffusion constant of the spin packet.  First of all
we will analyze the paramagnetic regime in which
$n_\uparrow=n_\downarrow$.  In Fig.\ref{fig8} we plot the ratio
$D_s/D_{ni}$, where $D_s$ represents the fully interacting calculation
according to Eq.~(\ref{dsfinal}) and $D_{ni}$ is the non-interacting
diffusion constant, as in \cite{Flatte}.  The figure shows results for
n-doped GaAs, at three different temperatures and in a range of densities
that is relevant to the experiments of Ref. \cite{Awscha5}.  We see that
the interaction correction is quite significant, and, in the paramagnetic
regime,
 reduces the value of the diffusion constant as expected, i.e.
$D_s/D_{ni}<1$ always.  The solid lines correspond to the calculations
performed at a temperature $T=300K$, the dashed lines to $T=20K$ and the
dot lines to $T=1.6K$.  The curves marked with "SD" correspond to the case
in which interactions in $D_s$ are taken into account only through the
spin Coulomb drag effect.  The figure shows clearly that at low
temperatures the most important many-body
 contribution to the diffusion is due to the softening of the spin
stiffness, while,
 already at $20K$ the spin drag contribution becomes relevant, to
represent most of the interaction effects at room temperature.  We see
that $D_s/D_{ni}\to 1$ for high densities. This is due to the enhancement
of screening in this regime, so that the particles tend to behave as
non-interacting ones. In the high density regime  $S\approx S_{ni}$
as can be seen in Fig.~\ref{fig6}.  If the temperature is high enough,
 the spin drag still reduces the diffusion constant by a sizable amount, but
eventually, even this contribution
disappears  with increasing  density,  and $D\to D_{ni}$.  

At low density, the system enters the non-degenerate regime, so that both
$D\to D_c$ and $D_{ni}\to D_c$, where $D_c = \mu_s k_BT/e$ is the classical
non-interacting diffusion constant. This limit will be discussed in greater detail
below.

In Fig.~\ref{fig9} we plot $D_s/D_c$ for the same parameters of
Fig.~\ref{fig8}. The solid lines represent the fully interacting diffusion
constant $D_s$
 while the dashed lines represent the corresponding non-interacting
approximation $D_{ni}$.  We see that, despite the significant reduction
due to the interaction correction,
 $D_s$ remains still considerably larger than $D_c$, consistent  with
experimental observations\cite{Awscha5}.

 As mentioned above, in the non-degenerate limit  $D/D_{c}$ approaches $1$. 
It is worthwhile to examine how this limit is approached.  In the
non-interacting theory \cite{Flatte} the non-degenerate
 limit is approached from above because the leading correction to $D_c$ is
due to
 the spin stiffness and comes from the quantum kinetic energy, i.e. it is
positive.  In the interacting theory there is an additional exchange
correction to $S$, which is negative, and competes with the quantum
kinetic one (see Sec.  \ref{nondegS}).  However,  the
leading correction to $D_c$ does not come from the spin stiffness term,
but from the spin Coulomb drag, and it is always {\it negative} (see
Eq.~(\ref{dsfinal})).  In fact, for $T>>T_F(n)$, $\rho_{\uparrow
\downarrow}/ \rho_D \sim [n/ (k_BT)^{3/2}] \ln (n/ (k_BT)^{2})$ (see
Eq.~(\ref{nondegrho2})), and the logarithmic term dominates over corrections
entering the spin stiffness in both the $n \to 0$ and $T\to\infty$ limits.
Thus, due to interactions, $D_s /D_c \to 1$ from {\it below} always.  This is
evident in
 Fig.\ref{fig8} where, for a fixed temperature, $D_s/D_c$ becomes negative
as the density decreases under a certain threshold, while for a fixed
density the ratio becomes negative when shifting to curves calculated at
higher temperature.

We want to stress that $D_s$ also displays a marked dependence on the
sample mobility that affects the diffusion constant through Drude
resistivity $\rho_D$. The higher the mobility, the more important becomes
the factor containing the spin transresistivity $1/(1-\rho_{\uparrow
\downarrow}/\rho_D)$ (see Eq.~(\ref{dsfinal})).  As we already underlined
in Sec. \ref{allT} the ratio $\rho_{\uparrow \downarrow}/\rho_D$ can
become very relevant and even greater than 1 (see Fig.~\ref{fig3} and
\ref{fig4}).  The diffusion constant in these cases is then regulated by
the spin drag effect that cannot be neglected.

Perhaps the most interesting feature of Eq.~(\ref{dsfinal}) is the
possibility of a large variation in $D_s$ when the electron gas undergoes
a ferromagnetic transition.  From the curves in Fig.\ref{fig6} and Fig.
\ref{fig10} we see that $S$ and $D_s$ vanish at the transition temperature
and increase sharply as the system settles in the fully polarized state.
 In the case of intrinsic ferromagnetism, the critical behavior of $D_s$
is completely due to Coulomb interactions among carriers.

As can be seen from Fig.~\ref{fig10}, $D_s /D_c\to 1$ as the system fully
polarizes. In this limit $n_\downarrow\to 0$, $\rho_{\downarrow\uparrow}/
\rho_D \to 0$, and $S/S_c \to 1$, as demonstrated in Sec. \ref{mixlimS},
so that
 $D_s$ reduces to the diffusion constant of carriers of minority
orientation (which are non-degenerate), i.e. to the classical value
$D_c=k_BT\mu/e$.

Unfortunately, in an ordinary electron liquid, the ferromagnetic
transition is predicted to occur only at extremely low densities.  There
is, however, an interesting variant:  semiconductors doped with {\it
magnetic} impurities (for example Mn)  can undergo a ferromagnetic
transition at rather high carrier densities \cite{Ohno5,Konig}, $n \sim
10^{20} cm^{-3}$ for (Ga,Mn)As, and temperatures $T<T_c \sim 110 K$
\cite{Ohno5}.  Our theory on the dynamics of a spin packet can be extended
to these systems with similar results \cite{EPL}.  This  extension will not  be
pursued here.

\section{Conclusions} In this paper, we have tried to demonstrate  the
importance of many-body effects in spin polarized transport.  We have
discussed in detail the spin Coulomb drag effect, an intrinsic source of friction
in spin transport that can limit spin currents even in the purest materials.
We have worked out  the behavior of the spin transresistivity
$\rho_{\uparrow\downarrow}$
 in  different physical regimes and shown that
 it ranges from $10^{-8}\Omega cm$ in metals to
$10^{-3}-10^{-2}\Omega cm$ in semiconductors.  Moreover, the ratio
$\rho_{\uparrow\downarrow}/\rho_D$, which is only a few percents in
metals,  becomes comparable to, or  even {\it larger than unity} in
semiconductors.  We hope that an experimental group will soon
take up the challenge of measuring the spin transresistivity, for example through
the experiment we suggest, in order to confirm the theory.

We have also demonstrated the importance of including
Coulomb interactions in a {\it quantitative} theory of spin diffusion, and shown
that a measure of $D_s$ for a unipolar spin packet would be a sensitive
probe of many-body effects such as the spin-Coulomb drag and the Coulomb
enhancement of the spin susceptibility.    

Finally we have studied  the behavior of $D_s$ at and below a
spontaneous ferromagnetic ordering transition and found that $D_s$
exhibits a critical behavior.

\section {Acknowledgements}  We  gratefully acknowledge support from  
NSF grant No. DMR-0074959.

\begin{figure}
  \caption{Spin transresistivity $\rho_{\uparrow\downarrow}$ as a function
of temperature (rescaled by $T_F$) for GaAs parameters ($m^*=0.067m_e$,
$\epsilon=12$, $\mu=3\times 10^3 cm^2/Vs$). Each curve corresponds to a
different density: $n_1=1.5\times 10^{16}cm^{-3}$, $n_2=1.5\times
10^{17}cm^{-3}$ $n_3=1.5\times 10^{18}cm^{-3}$.  Inset: comparison between
$\rho_{\uparrow\downarrow}$ and its analytical approximation in the
non-degenerate regime vs temperature (rescaled by $T_F$) for $n=n_2$.  }
  \label{fig2} \end{figure}

\begin{figure}
  \caption{Upper panel:  $\rho_{\uparrow\downarrow}/\rho_D$ as a function
of temperature for GaAs parameters ($m^*=0.067m_e$, $\epsilon=12$,
$n=1.5\times 10^{17}cm^{-3}$). Each curve corresponds to a different
mobility: $A=10^2 cm^2/Vs$, $B=10^3 cm^2/Vs$, $C=3\times 10^3 cm^2/Vs$
$D=10^4 cm^2/Vs$, as labelled. Lower panel: same as upper panel but for
(Ga,Mn)As parameters ($m^*=0.5m_e$, $\epsilon=12$, $n=1.2\times 10^{19}$).
  }
  \label{fig3} \end{figure}

\begin{figure}
  \caption{  $\rho_{\uparrow\downarrow}/\rho_D$ as a function
of carrier density for GaAs parameters ($m^*=0.067m_e$, $\epsilon=12$).
The solid curves are calculated at the temperature $T=300K$, while the
dashed curves at $T=20K$.  For each temperature two different mobilities
are considered,
 $C=3\times 10^3 cm^2/Vs$ and $D=10^4 cm^2/Vs$, as labelled.  }
  \label{fig4} \end{figure}

\begin{figure}
  \caption{Temperature and density dependence of
$\rho_{\uparrow\downarrow}$ in the degenerate regime.
The top line corresponds to $r_s=7$.  The electron-gas parameter is
decremented by 1 starting from the top.  }
  \label{fig5} \end{figure}

\begin{figure}
  \caption{Spin stiffness $S$, rescaled by its non-interacting
approximation $S_{ni}$ vs $T/T_F$. The carrier density is $n=4.2\times
10^{11}cm^{-3}$ for the lower curve and increases by a factor $10$ for
each line starting from the bottom.  The cusps correspond to the onset of
ferromagnetism.  }
  \label{fig6} \end{figure}

\begin{figure}
  \caption{(a) Experimental setup to detect the spin Coulomb drag effect:
the voltage $\Delta V$ is applied between two parallel half-metallic
ferromagnets (injector (inj.) and receiver (rec.)) that sandwich a
paramagnet (P). The voltage $\Delta V_D$ is detected using two
ferromagnetic electrodes (d)
 similar to the injector and the receiver, but polarized in the {\it
opposite} direction.  (b) Schematic band-structure of injector, receiver,
detectors  and paramagnet P. 
(c) Schematic behavior of the chemical ($\xi_\downarrow$) and
electro-chemical  potentials $\psi_\alpha(r)$ 
(see Eq.~\ref{electrochemicalpotential}). For this setup
$\xi_\uparrow$ is basically constant (not shown in the figure). 
} \label{fig7} \end{figure}

\begin{figure} \caption{ The interacting diffusion constant of a
spin-packet $D_s$ rescaled by its non-interacting approximation $D_{ni}$
vs
 density for different temperatures: solid lines correspond to $T=300K$,
dashed lines to $T=20K$ and dotted lines to $T=1.6K$.  For each
temperature, we plot also  the curve
 obtained considering 
 interactions only through the spin Coulomb drag effect (labeled as
SD).  In all the calculations the dielectric constant of the semiconductor
is $\epsilon =12$ and the mobility is $\mu = 3\times 10^3 cm^2/Vs$.  }
\label{fig8} \end{figure}

\begin{figure} \caption{ The interacting (solid line) and non-interacting
(dashed line)  diffusion constant of a spin-packet rescaled by its
classical non-interacting approximation $D_{c}$ vs
 density for different temperatures.
 In all the calculations the dielectric constant of the semiconductor is
$\epsilon =12$ and the mobility is $\mu = 3\times 10^3 cm^2/Vs$.  }
\label{fig9} \end{figure}

\begin{figure} \caption{ The diffusion constant of a spin packet
rescaled by its classical non-interacting approximation $D_{c}$ vs
temperature in a low-density electron gas with no magnetic impurities.
  } \label{fig10} \end{figure}

\newpage \begin{figure} 
\psfig{figure=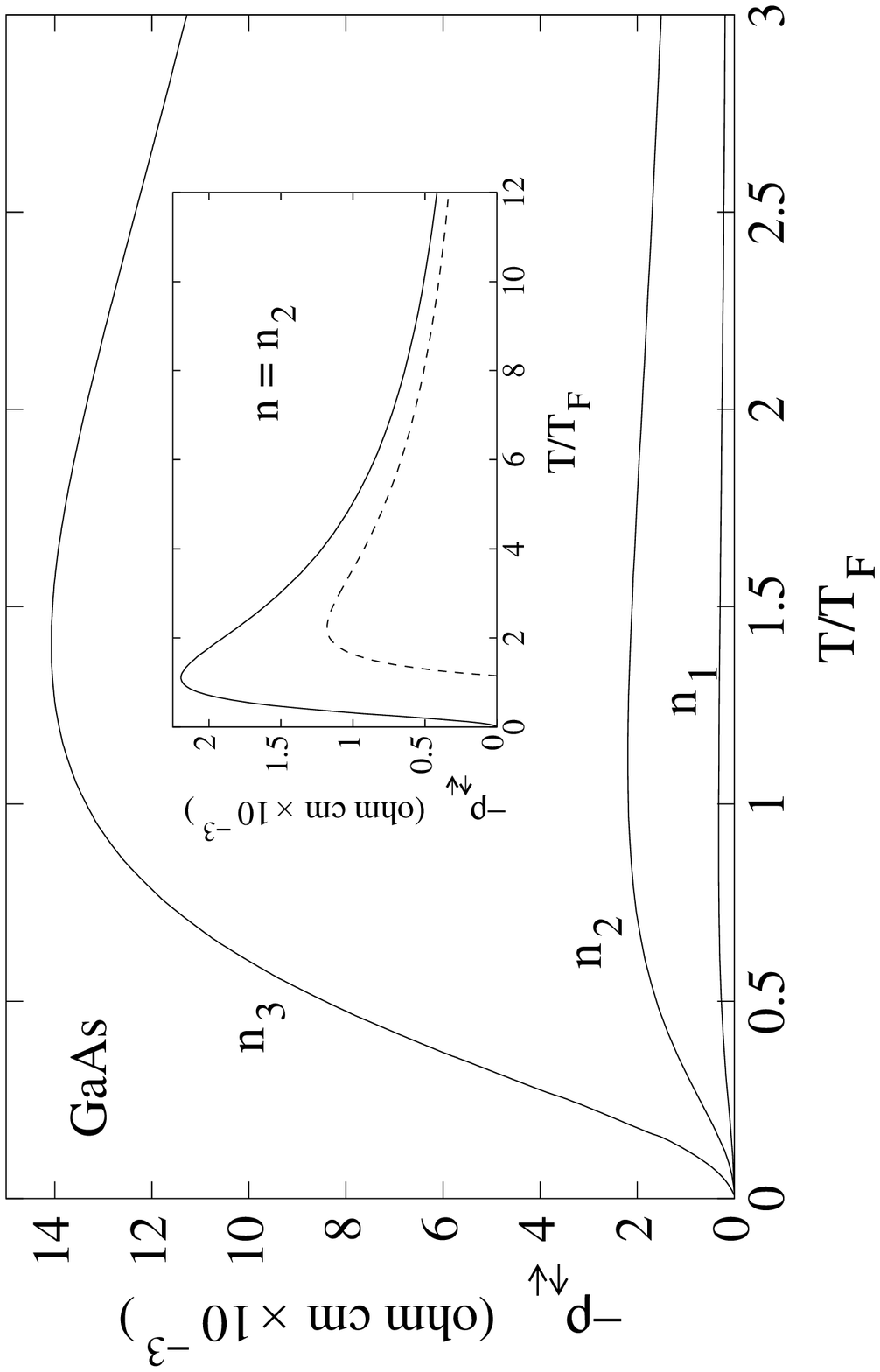,width=1.00\columnwidth,angle=0}
\end{figure} \Large{Fig. \ref{fig2}}

\newpage \begin{figure} 
\psfig{figure=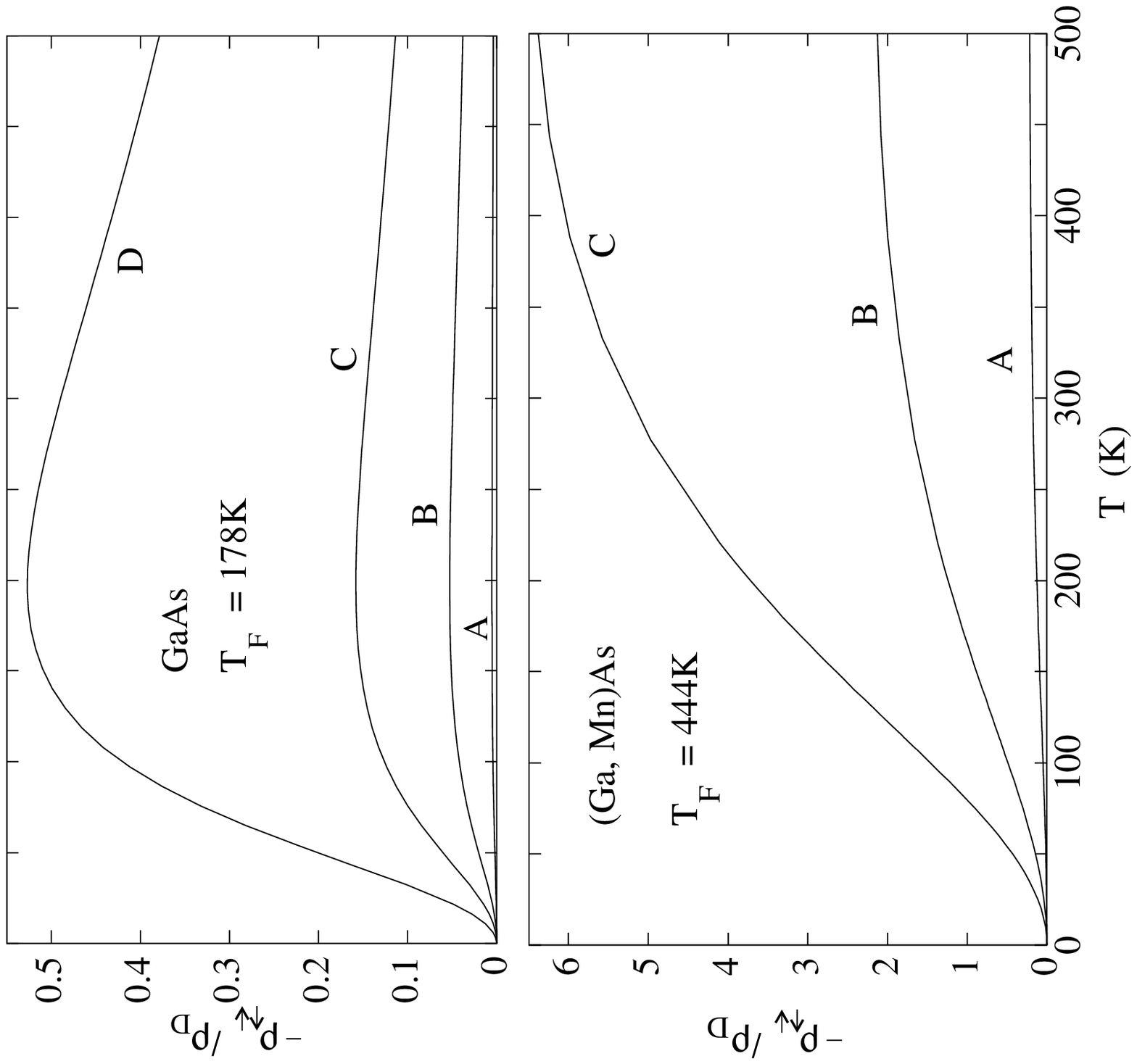,width=1.0\columnwidth,angle=0}
\end{figure} \Large{Fig. \ref{fig3}}

\newpage \begin{figure} 
\psfig{figure=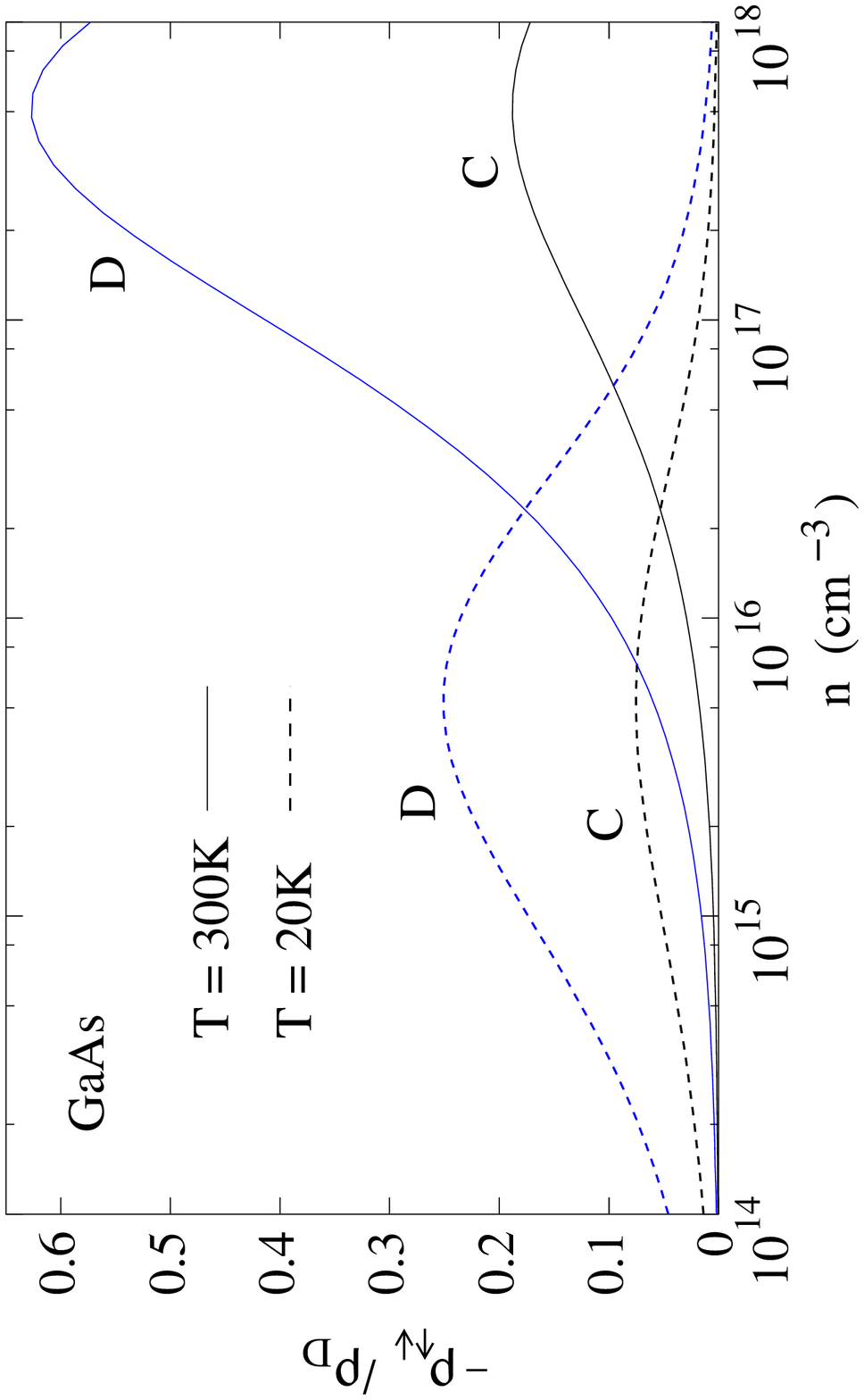,width=1.0\columnwidth,angle=0}
\end{figure} \Large{Fig. \ref{fig4}}

\newpage \begin{figure} 
\psfig{figure=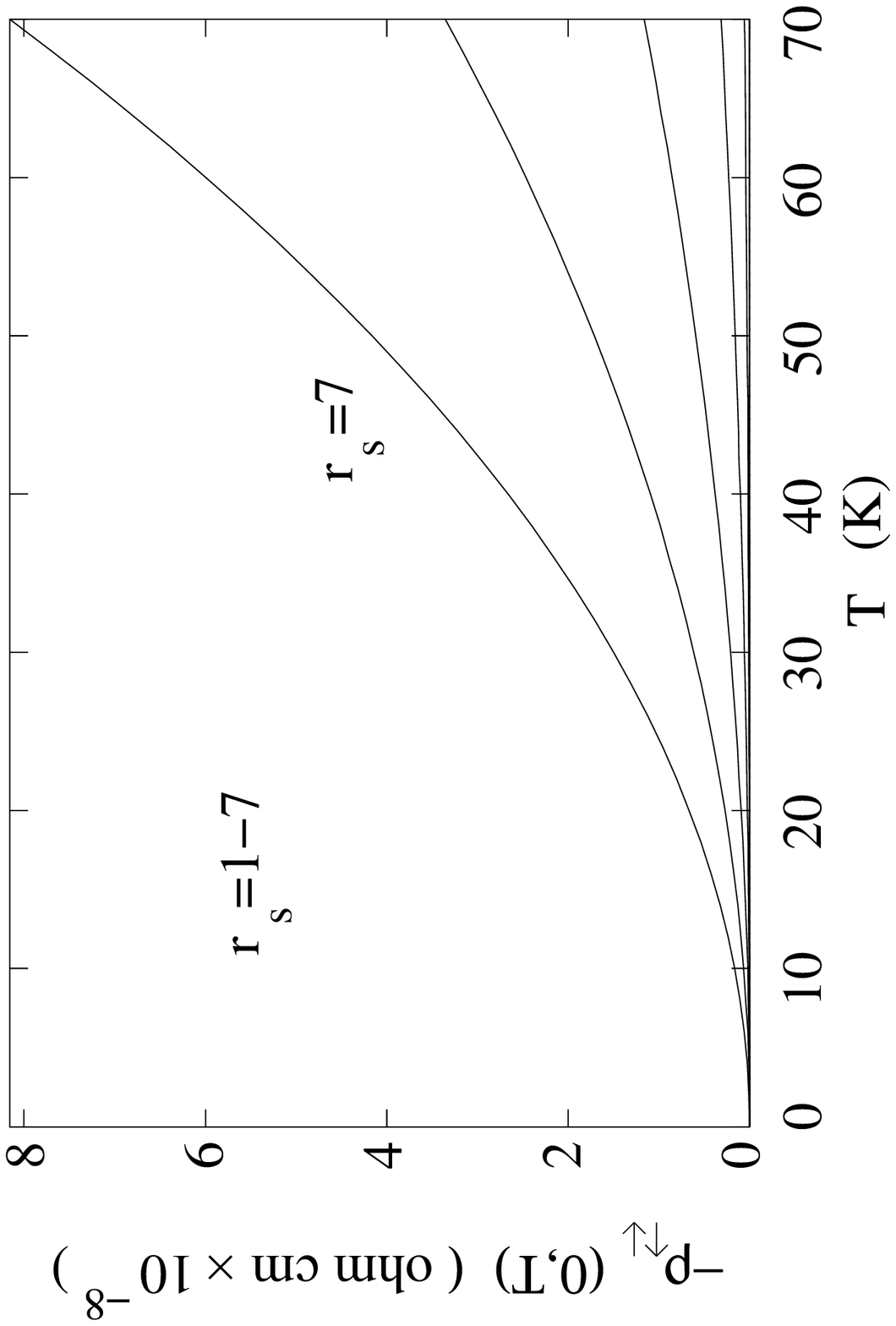,width=1.0\columnwidth,angle=0}
\end{figure} \Large{Fig. \ref{fig5}}

\newpage \begin{figure} 
\psfig{figure=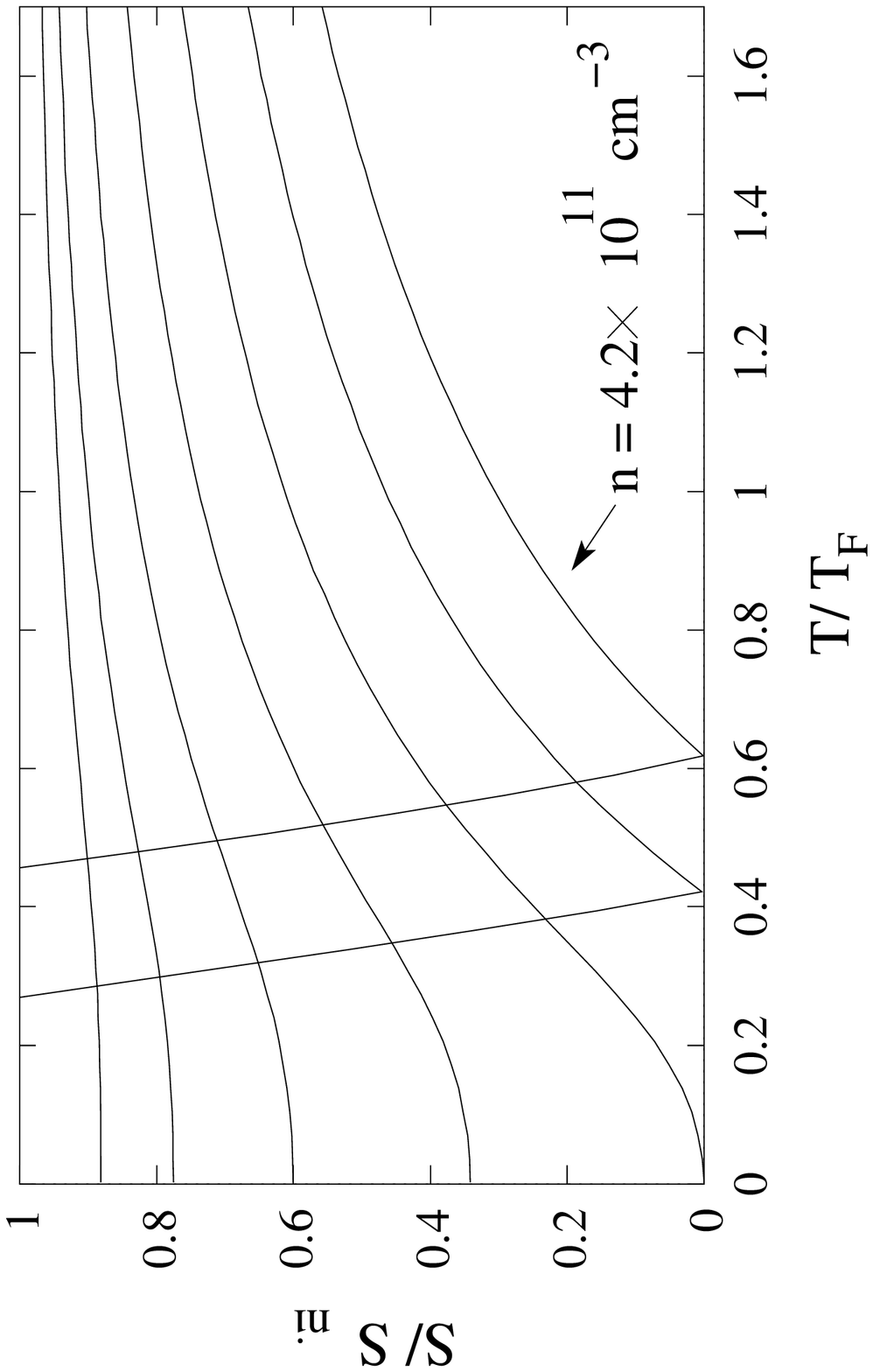,width=1.0\columnwidth,angle=0}
\end{figure} \Large{Fig. \ref{fig6}}

\newpage \begin{figure} 
\psfig{figure=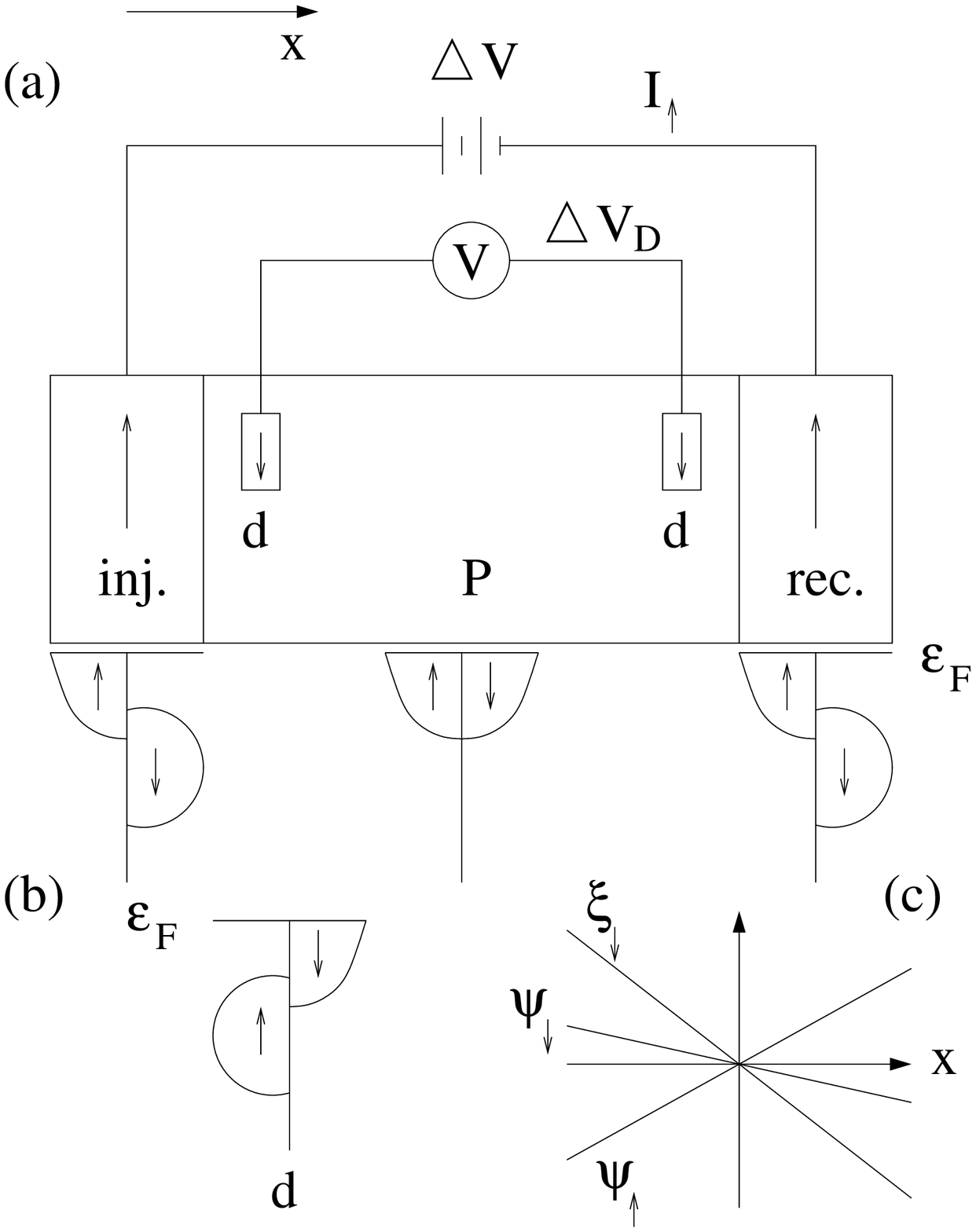,width=1.0\columnwidth,angle=0}
\end{figure} \Large{Fig. \ref{fig7}}

\newpage \begin{figure} 
\psfig{figure=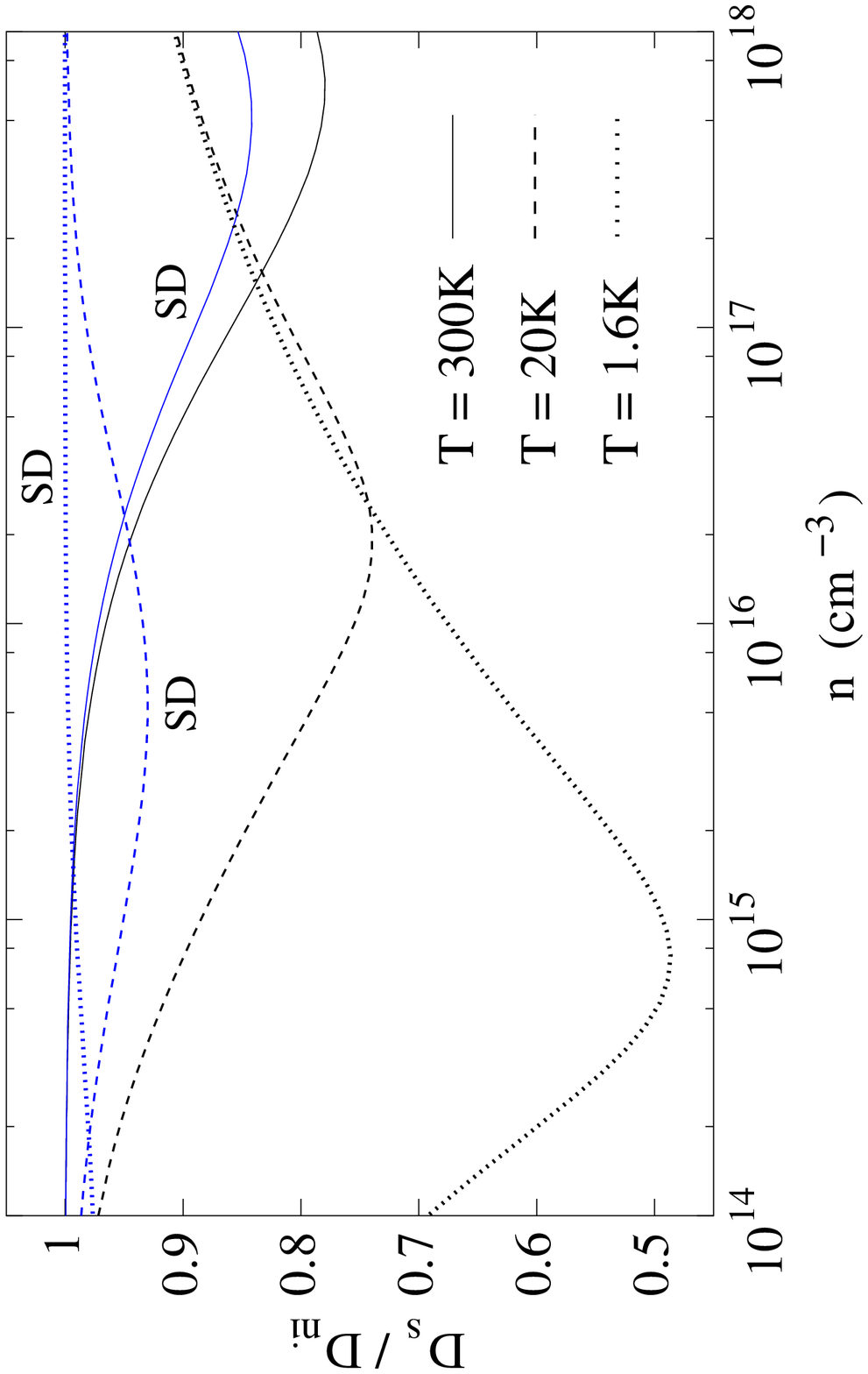,width=1.0\columnwidth,angle=0}
\end{figure} \Large{Fig. \ref{fig8}}

\newpage \begin{figure} 
\psfig{figure=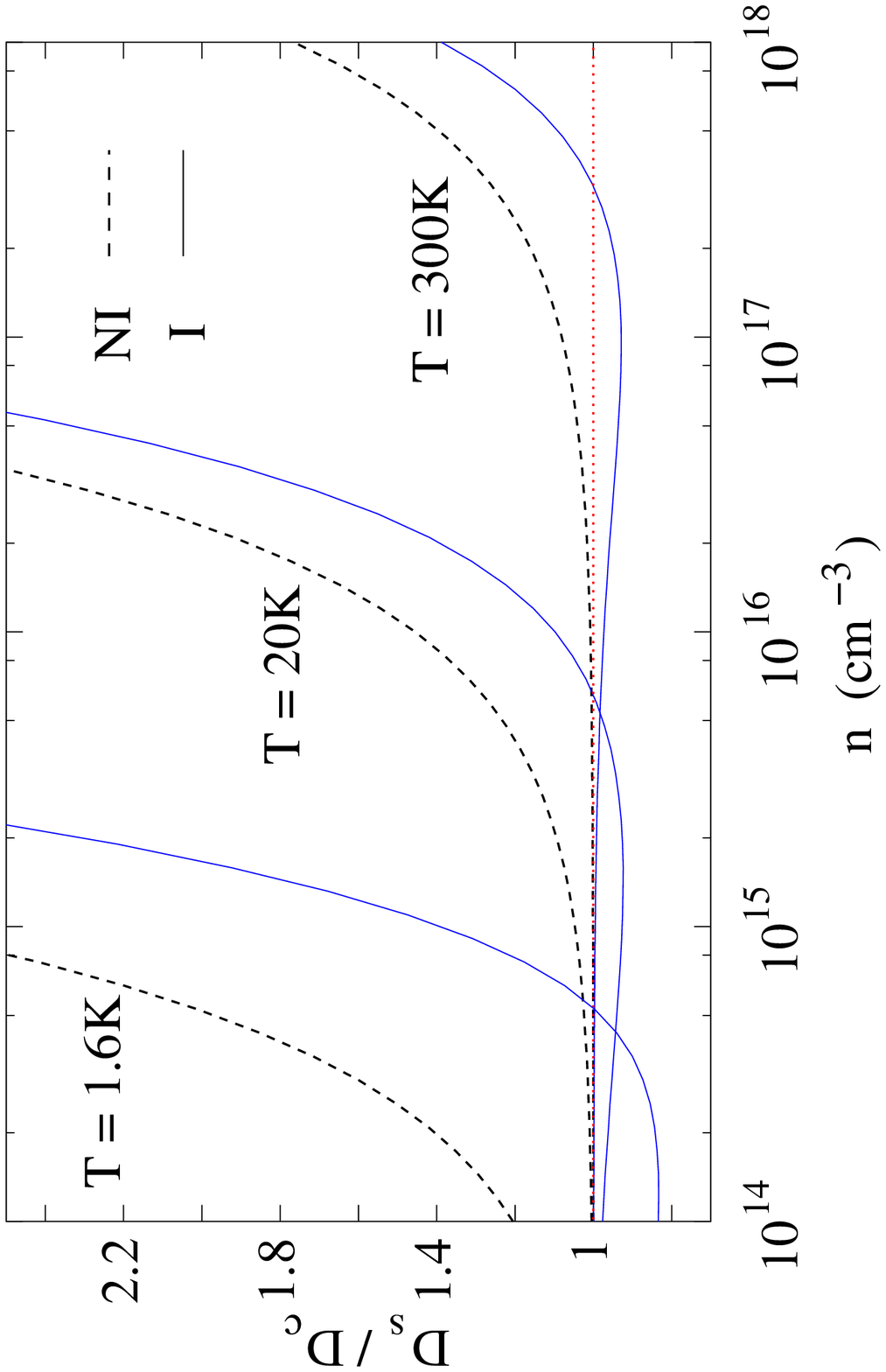,width=1.0\columnwidth,angle=0}
\end{figure} \Large{Fig. \ref{fig9}}

\newpage 
\begin{figure} 
\psfig{figure=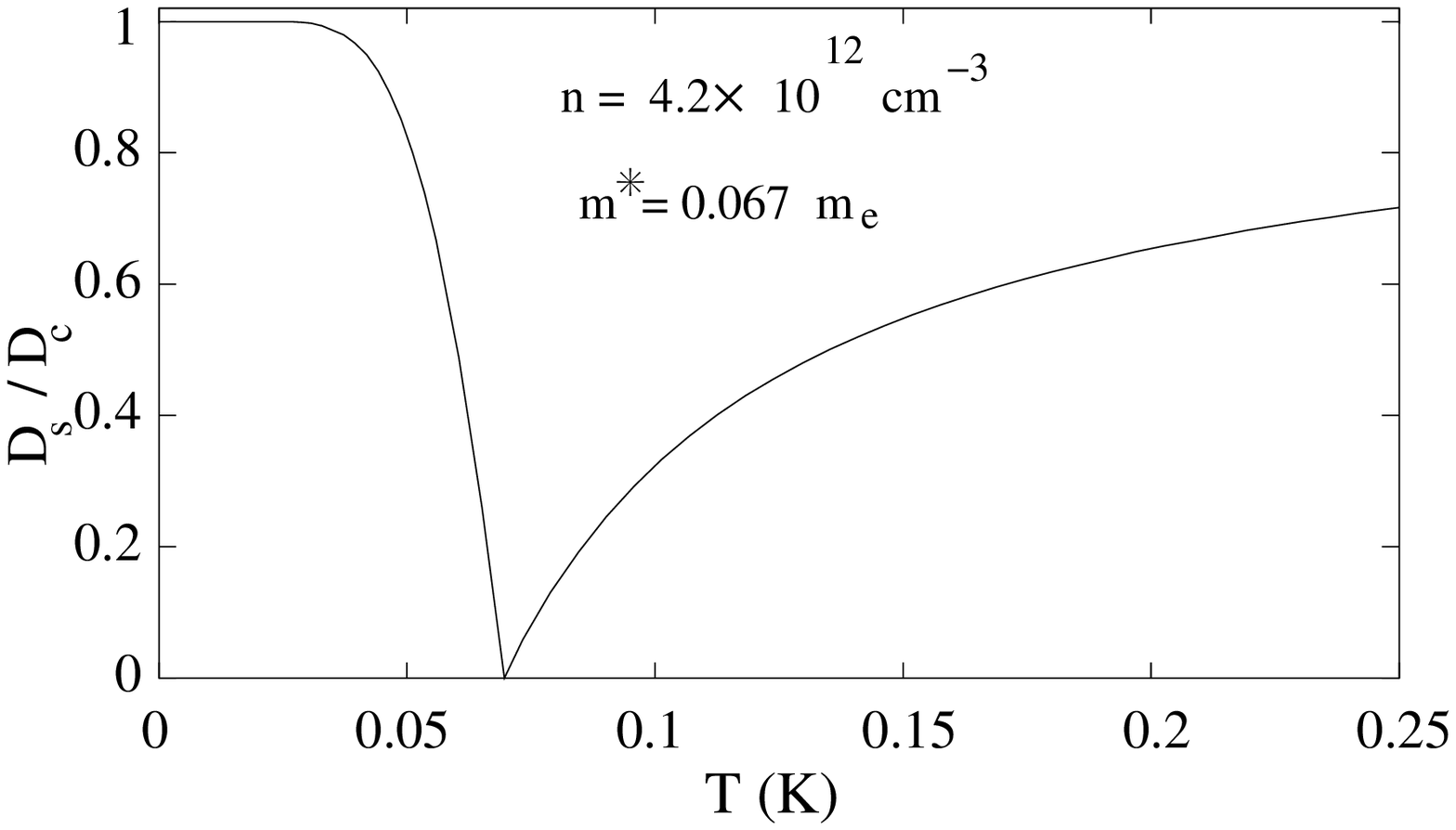,width=1.4\columnwidth,angle=90}
\end{figure} \Large{Fig. \ref{fig10}}


\begin{thebibliography}{aaaaaaaaaaaaaaaaaaaa}
\bibitem{footnote0}For simplicity, we limit ourselves in this paper to
either the
paramagnetic case or the ferromagnetic case with collinear spins.
\bibitem{footnote1} We assume that impurities have a negligible effect on
the thermodynamic functions of the electron gas.
\bibitem{footnote2} In the weak coupling limit, the  calculations reviewed in
this section are completely equivalent to an evaluation of  $\gamma$ 
according
to the Boltzmann-equation formula (\ref{taucoulomb}).  More generally,
however,  the Kubo formalism allows us to take into account some strong
coupling effects, such as the finite spectral width of the Landau
quasiparticles.
\bibitem{footnote3} In the spirit of the linear response approximation
these constants should be evaluated at the underlying equilibrium density
of the electron gas.
\bibitem{footnote4}  Ideally one would like to use an injector that is closely
matched in impedance to the paramagnet, so that all the injected current
goes into
an average drift velocity of the up-spin component and none is ``lost"  in
the form of
a diffusion current.
\end{thebibliography}
\end{document}